\documentclass[12pt]{article}
\usepackage{graphicx}
\def\ffrac#1#2{\textstyle{#1\over#2}\displaystyle}

\begin{document}
\pagestyle{myheadings}
\parindent 0mm
\parskip 6pt
\markright{SLE for Theoretical Physicists}

\title{SLE for Theoretical Physicists}

\author{John Cardy\\
Rudolf Peierls Centre for Theoretical Physics\\
         1 Keble Road, Oxford OX1 3NP,
         U.K.\footnote{Address for correspondence.}\\
and All Souls College, Oxford.}
\date{May 2005}
\maketitle
\begin{abstract}
This article provides an introduction to
Schramm(stochastic)-Loewner evolution (SLE) and to its connection
with conformal field theory, from the point of view of its
application to two-dimensional critical behaviour. The emphasis is
on the conceptual ideas rather than rigorous proofs.
\end{abstract}
\newpage

\section{Introduction}
\subsection{Historical overview}
The study of critical phenomena has long been a breeding ground
for new ideas in theoretical physics. Such behaviour is
characterised by a diverging correlation length and cannot easily
be approximated by considering small systems with only a few
degrees of freedom. Initially, it appeared that the analytic study
of such problems was a hopeless task, although self-consistent
approaches such as mean field theory were often successful in
providing a semi-quantitative description.

Following Onsager's calculation of the free energy of the square
lattice Ising model in 1944, steady progress was made in the exact
solution of an ever-increasing number of lattice models in two
dimensions. While many of these are physically relevant, and the
techniques used have spawned numerous spin-offs in the theory of
integrable systems, it is fair to say that these methods have not
cast much light on the general nature of the critical state. In
addition, virtually no progress has been made from this direction
in finding analytic solutions to such simple and important
problems as percolation.

An important breakthrough occurred in the late 1960's, with the
development of renormalisation group (RG) ideas by Wilson and
others. The fundamental realisation was that, in the scaling limit
where both the correlation length and all other macroscopic length
scales are much larger than that of the microscopic interactions,
classical critical systems are equivalent to renormalisable
quantum field theories in euclidean space-time. Since there is
often only a finite or a denumerable set of such field theories
with given symmetries, at a stroke this explained the observed
phenomenon of universality: systems with very different
constituents and microscopic interactions nevertheless exhibit the
same critical behaviour in the scaling limit. This single idea has
led to a remarkable unification of the theoretical bases of
particle physics, statistical mechanics and condensed matter
theory, and has led to extensive cross-fertilisation between these
disciplines. These days, a typical paper using the ideas and
methods of quantum field theory is as likely to appear in a
condensed matter physics journal as in a particle physics
publication (although there seems to be a considerable degree of
conservatism among the writers of field theory text books in
recognising this fact.)

Two important examples of this interdisciplinary flow were the
development of lattice gauge theories in particle physics, and the
application of conformal field theory (CFT), first developed as a
tool in string theory, to statistical mechanics and condensed
matter physics. As will be explained later, in two-dimensional
classical systems and quantum systems in 1+1 dimensions conformal
symmetry is extremely powerful, and has led to a cornucopia of new
exact results. Essentially, the RG programme of classifying all
suitable renormalisable quantum field theories in two dimensions
has been carried through to its conclusion in many cases,
providing exact expressions for critical exponents, correlation
functions, and other universal quantities. However the
geometrical, as opposed to the algebraic, aspects of conformal
symmetry are not apparent in this approach.

One minor but nevertheless theoretically influential prediction of
these methods was the conjectured crossing formula
\cite{JCcrossing} for the probability that, in critical
percolation, a cluster should exist which spans between two
disjoint segments of the boundary of some simply connected region
(a more detailed account of this problem will be given later.)
With this result, the simmering unease that mathematicians felt
about these methods came to the surface (see, for example, the
comments in \cite{Lang}.) What exactly are these renormalised
local operators whose correlation functions the field theorists so
happily manipulate, according to rules that sometimes seem to be a
matter of cultural convention rather than any rigorous logic? What
does conformal symmetry really mean? Exactly which object is
conformally invariant? And so on. Aside from these deep concerns,
there was perhaps also the territorial feeling that percolation
theory, in particular, is a branch of probability theory, and
should be understood from that point of view, not merely as a
by-product of quantum field theory.

Thus it was that a number of pure mathematicians, versed in the
methods of probability theory, stochastic analysis and conformal
mapping theory, attacked this problem. Instead of trying make
rigorous the notions of field theory about local operators, they
focused on the random curves which form the boundaries of clusters
on the lattice, and on what should be the properties of the
measure on such curves in the continuum limit as the lattice
spacing approaches zero. The idea of thinking about lattice models
this way was not new: in particular in the 1980s it led to the
very successful but non-rigorous Coulomb gas approach
\cite{Nienhuis} to two-dimensional critical behaviour, whose
results parallel and complement those of CFT. However, the new
approach focused on the properties of a single such curve,
conditioned to start at the boundary of the domain, in the
background of all the others. This leads to a very specific and
physically clear notion of conformal invariance. Moreover, it was
shown by Loewner\cite{Loewner} in the 1920s that any such curve in
the plane which does not cross itself can be described by a
dynamical process called Loewner evolution, in which the curve is
imagined to be grown in a continuous fashion. Instead of
describing this process directly, Loewner considered the evolution
of the analytic function which conformally maps the region outside
the curve into a standard domain. This evolution, and therefore
the curve itself, turns out to be completely determined by a real
continuous function $a_t$. For random curves, $a_t$ itself is
random. (The notation $a_t$ is used rather than $a(t)$ to conform
to standard usage in the case when it is a stochastic variable.)
Schramm\cite{Schramm} argued that, if the measure on the curve is
to be conformally invariant in the precise sense referred to
above, the only possibility is that $a_t$ be one-dimensional
Brownian motion, with only a single parameter left undetermined,
namely the diffusion constant $\kappa$. This leads to stochastic-,
or Schramm-, Loewner evolution (SLE). (In the original papers by
Schramm et al. the term `stochastic' was used. However in the
subsequent literature the `S' has often been taken to stand for
Schramm in recognition of his contribution.) It should apply to
any critical statistical mechanics model in which it is possible
to identify these non-crossing paths on the lattice, as long as
their continuum limits obey the underlying conformal invariance
property. For only a few cases, including percolation, has it been
proved that this property holds, but it is believed to be true for
suitably defined curves in a whole class of systems known as
O$(n)$ models. Special cases, apart from percolation, include the
Ising model, Potts models, the XY model, and self-avoiding walks.
They each correspond to a particular choice of $\kappa$.

Starting from the assumption that SLE describes such a single
curve in one of these systems, many properties, such as the values
of many of the critical exponents, as well as the crossing formula
mentioned above, have been rigorously derived in a brilliant
series of papers by Lawler, Schramm and Werner (LSW)\cite{LSW}.
Together with Smirnov's proof \cite{Smirnov} of the conformal
invariance property for the continuum limit of site percolation on
the triangular lattice, they give a \em rigorous \em derivation
\cite{SmirnovWerner} of the values of the critical exponents for
two-dimensional percolation. This represents a paradigm shift in
rigorous statistical mechanics, in that results are now being
derived directly in the continuum for models for which the
traditional lattice methods have, so far, failed.

However, from the point of view of theoretical physics, these
advances are  perhaps not so important for being rigorous, as for
the new light they throw on the nature of the critical state, and
on conformal field theory. In the CFT of the O$(n)$ model, the
point where a random curve hits the boundary corresponds to the
insertion of a local operator which has a particularly simple
property: its correlation functions satisfy linear second-order
differential equations\cite{JCsurf}. These equations turn out to
be directly related to the Fokker-Planck type equations one gets
from the Brownian process which drives SLE. Thus there is a close
connection, at least at an operational level, between CFT and SLE.
This has been made explicit in a series of papers by Bauer and
Bernard\cite{BB} (see also \cite{FriedrichWerner}.) Other
fundamental concepts of CFT, such as the central charge $c$, have
their equivalence in SLE. This is a rapidly advancing subject, and
some of the more recent directions will be mentioned in the
concluding section of this article.
\subsection{Aims of this article}
The original papers on SLE are mostly both long and difficult,
using, moreover, concepts and methods foreign to most theoretical
physicists. There are reviews, in particular those by
Werner\cite{Wreview} and by Lawler\cite{Lreview} which cover much
of the important material in the original papers. These are
however written for mathematicians. A more recent review by Kager
and Nienhuis\cite{KNreview} describes some of the mathematics in
those papers in way more accessible to theoretical physicists, and
should be essential reading for any reader who wants then to
tackle the mathematical literature. A complete bibliography up to
2003 appears in \cite{KadGruz}.

However, the aims of the present article are more modest. First,
it does not claim to be a thorough review, but rather a
semi-pedagogical introduction. In fact some of the material,
presenting some of the existing results from a slightly different,
and hopefully clearer, point of view, has not appeared before in
print. The article is directed at the theoretical physicist
familiar with the basic concepts of quantum field theory and
critical behaviour at the level of a standard graduate textbook,
and with a theoretical physicist's knowledge of conformal mappings
and stochastic processes. It is not the purpose to prove anything,
but rather to describe the concepts and methods of SLE, to relate
them to other ideas in theoretical physics, in particular CFT, and
to illustrate them with a few simple computations, which, however,
will be presented in a thoroughly non-rigorous manner. Thus, this
review is most definitely not for mathematicians interested in
learning about SLE, who will no doubt cringe at the lack of
preciseness in some of the arguments and perhaps be puzzled by the
particular choice of material. The notation used will be that of
theoretical physics, for example $\langle\cdots\rangle$ for
expectation value, and so will the terminology. The word
`martingale' has just made its only appearance. Perhaps the
largest omission is any account of the central arguments of
LSW\cite{LSW} which relate SLE to various aspects of Brownian
motion and thus allow for the direct computation of many critical
exponents. These methods are in fact related to two-dimensional
quantum gravity, whose role in this is already the subject of a
recent long article by Duplantier\cite{Dupreview}.

\subsubsection*{Acknowledgments.}
One may go only so far in learning a subject like this by reading
the original literature, and much of my knowledge, such as it is,
comes from seminars by, and informal discussions with, a large
number of people. To them I give thanks, as well as apologies if I
have occasionally misrepresented or oversimplified their ideas:
M.~Aizenman, M.~Bauer, R.~Bauer, V.~Beffara, J.~Dub\'edat,
D.~Bernard, M.~den Nijs, B.~Doyon, B.~Duplantier, R.~Friedrich,
I.~Gruzberg, M.~Hastings, J.~Jacobsen, L.~Kadanoff, W.~Kager,
R.~Kenyon, H.~Kesten, P.~Kleban, J.~Kondev, R.~Langlands,
G.~Lawler, B.~Nienhuis, S.~Rohde, Y.~Saint-Aubin, H.~Saleur,
O.~Schramm, S.~Sheffield, S.~Smirnov, W.~Werner, D.~Wilson. I also
thank B.~Doyon and V.~Riva for comments on the manuscript. This
work was begun while the author was a Member of the Institute for
Advanced Study, supported by the Ellentuck Fund.
\newpage
\section{Random curves and lattice models.}
\label{sec:lattice}
\subsection{The Ising and percolation models}
\label{sec:ising} In this section, we introduce the lattice models
which can be interpreted in terms of random non-intersecting paths
on the lattice whose continuum limit will be described by SLE.

\begin{figure}[h]
\centering
\includegraphics[width=8cm]{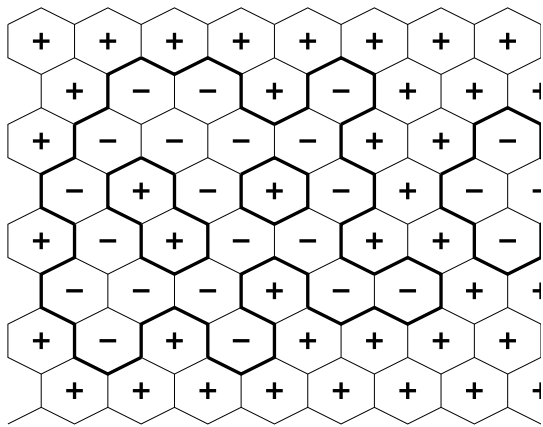}
\caption{\label{fig:hex}\small
Ising model on the honeycomb lattice, with loops corresponding to a
term in the expansion of (\ref{ZI}). Alternatively, these may be
thought of as domain walls of an Ising model on the dual triangular
lattice.}
\end{figure}

The prototype is the Ising model. It is most easily realised on a
honeycomb lattice (see Fig.~\ref{fig:hex}).
At each site $r$ is an Ising `spin' $s(r)$ which takes the
values $\pm1$. The partition function is
\begin{equation}
\label{ZI} Z_{\rm Ising}={\rm Tr}\,e^{\beta J\sum_{rr'}s(r)s(r')}
\propto{\rm Tr}\,\prod_{rr'}\left(1+xs(r)s(r')\right)\,,
\end{equation}
where $x=\tanh\beta J$, and the sum and product are over all edges
joining nearest neighbour pairs of sites. The trace operation is
defined as ${\rm Tr}=\prod_r\big(\frac12\sum_{s(r)}\big)$, so that
${\rm Tr}\,s(r)^n=1$ if $n$ is even, and $0$ if it is odd.

At high temperatures ($\beta J\ll 1$) the spins are disordered,
and their correlations decay exponentially fast, while at low
temperatures ($\beta J\gg1$) there is long-range order: if the
spins on the boundary are fixed say, to the value $+1$, then
$\langle s(r)\rangle\not=0$ even in the infinite volume limit. In
between, there is a critical point. The conventional approach to
the Ising model focuses on the behaviour of the correlation
functions of the spins. In the scaling limit, they become local
operators in a quantum field theory (QFT). Their correlations are
power-law behaved at the critical point, which corresponds to a
massless QFT, that is a conformal field theory (CFT). From this
point of view (as well as exact lattice calculations) it is found
that correlation functions like $\langle s(r_1)s(r_2)\rangle$
decay at large separations according to power laws
$|r_1-r_2|^{-2x}$: one of the aims of the theory is to obtain the
values of the exponents $x$ as well as to compute, for example,
correlators depending on more than two points.

However, there is an alternative way of thinking about the
partition function (\ref{ZI}), as follows: imagine expanding out
the product to obtain $2^N$ terms, where $N$ is the total number
of edges. Each term may be represented by a subset of edges, or
graph $\cal G$, on the lattice, in which, if the term $xs(r)s(r')$
is chosen, the corresponding edge $(rr')$ is included in $\cal G$,
otherwise it is not. Each site $r$ has either 0, 1, 2 or 3 edges
in $\cal G$. The trace over $s(r)$ gives 1 if this number is even,
and 0 if it is odd. Each surviving graph is then the union of
non-intersecting closed loops (see Fig.~\ref{fig:hex}). In
addition, there can be open paths beginning and ending at a
boundary. For the time being, we suppress these by imposing `free'
boundary conditions, summing over the spins on the boundary. The
partition function is then
\begin{equation}
Z_{\rm Ising}=\sum_{\cal G}\,x^{\rm length}\,,
\end{equation}
where the length is the total of all the loops in $\cal G$.
When $x$ is small, the mean length of a single loop is small. The
critical point $x_c$ is signalled by a divergence of this quantity.
The low-temperature phase corresponds to $x>x_c$. While in this phase
the Ising spins are ordered, and their connected correlation functions\
decay exponentially, the loop gas is in fact still critical, in that, for
example, the probability that two points lie on the same loop has
a power-law dependence on their separation. This is the
\em dense \em phase.

The loops in $\cal G$ may be viewed in another way: as \em domain
walls \em for another Ising model on the dual lattice, which is a
triangular lattice whose sites $R$ lie at the centres of the
hexagons of the honeycomb lattice (see Fig.~\ref{fig:hex}). If the
corresponding interaction strength of this dual Ising model is
$(\beta J)^*$, then the Boltzmann weight for creating a segment of
domain wall is $e^{-2(\beta J)^*}$. This should be equated to
$x=\tanh(\beta J)$ above. Thus we see that the high-temperature
regime of the dual model corresponds to low temperature in the
original model, and vice versa. Infinite temperature in the dual
model ($(\beta J)^*=0$) means that the dual Ising spins are
independent random variables. If we colour each dual site with
$s(R)=+1$ black, and white if $s(R)=-1$, we have the problem of
\em site percolation \em on the triangular lattice, critical
because $p_c=\frac12$ for that problem. Thus the curves with $x=1$
correspond to percolation cluster boundaries. (In fact in the
scaling limit this is believed to be true throughout the dense
phase $x>x_c$.)

So far we have discussed only closed loops. Consider the spin-spin
correlation function
\begin{equation}
\langle s(r_1)s(r_2)\rangle= {{\rm
Tr}\,s(r_1)s(r_2)\prod_{rr'}\left(1+xs(r)s(r')\right) \over{\rm
Tr}\,\prod_{rr'}\left(1+xs(r)s(r')\right)}\,,
\end{equation}
where the sites $r_1$ and $r_2$ lie on the boundary. Expanding out
as before, we see that the surviving graphs in the numerator each
have a single edge coming into $r_1$ and $r_2$. There is therefore
a single open path $\gamma$ connecting these points on the
boundary (which does not intersect itself nor any of the closed
loops.) In terms of the dual variables, such a single open curve
may be realised by specifying the spins $s(R)$ on all the dual
sites on the boundary to be $+1$ on the part of the boundary
between $r_1$ and $r_2$ (going clockwise) and $-1$ on the
remainder. There is then a single domain wall connecting $r_1$ to
$r_2$. {\it SLE describes the continuum limit of such a curve
$\gamma$.}

Note that we could also choose $r_2$ to lie in the interior. The
continuum limit of such curves is then described by radial SLE
(Sec.~\ref{sec:radialsle}).

\subsubsection{Exploration process} \label{sec:exploration}
\begin{figure}
\centering
\includegraphics[width=8cm]{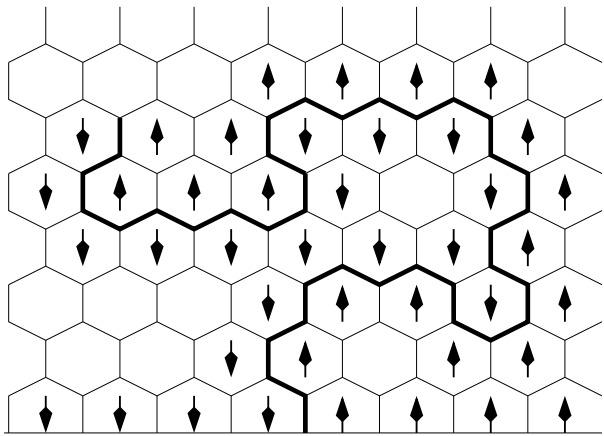}
\caption{\label{fig:exploration}\small
The exploration process for the Ising model. At each step the walk
turns L or R according to the value of the spin in front of it. The
relative probabilities are determined by the expectation value of this
spin given the fixed spins either side of the walk up to this time. The
walk never crosses itself and never gets trapped.}
\end{figure}

An important property of the ensemble of curves $\gamma$ on the
lattice is that, instead of generating a configuration of all the
$s(R)$ and then identifying the curve, it may be constructed
step-by-step as follows (see Fig.~\ref{fig:exploration}).
Starting
from $r_1$, at the next step it should turn R or L according to
whether the spin in front of it is $+1$ or $-1$. For independent
percolation, the probability of either event is $\frac12$, but for
$x<1$ it depends on the values of the spins on the boundary.
Proceeding like this, the curve will grow, with all the dual sites
on its immediate left taking the value $+1$, and those its right
the value $-1$. The relative probabilities of the path turning R
or L at a given step depend on the expectation value of the spin
on the site $R$ immediately in front of it, given the values of
the spins already determined, that is, given the path up to that
point. Thus the relative probabilities that the path turns R or L
are completely determined by the domain and the path up to that
point. This implies the crucial

{\bf Property 2.1} (lattice version). Let $\gamma_1$ be the part
of the total path covered after a certain number of steps. Then
the conditional probability distribution of the remaining part of
the curve, given $\gamma_1$, is the same as the unconditional
distribution of a whole curve, starting at the tip and ending at
$r_2$ in the domain ${\cal D}\setminus\gamma_1$.

In the Ising model, for example, if we already know part of the domain wall,
the rest of it can be considered as a complete domain
wall in a new region in which the left and right sides of the existing part
form part of the boundary.
This means the path is a history-dependent random walk. It can
be seen (Fig.~\ref{fig:exploration}) that when the growing tip
$\tau$
approaches an earlier section of the path, it must always turn away
from it: the tip never gets trapped.
There is always at least one path on the lattice from
the tip $\tau$ to the final point $r_2$.

\subsection{O$(n)$ model}\label{sec:Onmodel}
The loop gas picture of the Ising and percolation models may
simply be generalised by counting each closed loop with a fugacity
$n$:
\begin{equation}\label{On1}
Z_{{\rm O}(n)}=\sum_{\cal G}\,x^{\rm length}\,n^{\rm number\ of\
loops}\,.
\end{equation}
This is called the O$(n)$ model, for the reason that it gives the
partition function for $n$-component spins
${\bf s}(r)=(s_1(r),\ldots s_n(r))$
with
\begin{equation}\label{On2}
Z_{{\rm O}(n)}={\rm Tr}\,\prod_{rr'}\left(1+ x{\bf s}(r)\cdot{\bf
s}(r')\right)\,,
\end{equation}
where ${\rm Tr}\,s_a(r)s_b(r)=\delta_{ab}$. Following the same
procedure as before we obtain the same set of closed loops (and
open paths) except that, on summing over the last spin in each
closed loop, we get a factor $n$. The model is called O$(n)$
because of its symmetry under rotations of the spins. The version
(\ref{On2}) makes sense only when $n$ is a positive integer (and
note that the form of the partition function is different from
that of the conventional O$(n)$ model, where the second term is
exponentiated.) The form in (\ref{On1}) is valid for general
values of $n$, and it gives a probability measure on the loop gas
for real $n\geq0$. However, the dual picture is useful only for
$n=1$ and $n=2$ (see below.) As for the case $n=1$, there is a
critical value $x_c(n)$ at which the mean loop length diverges.
Beyond this, there is a dense phase.

Apart from $n=1$, other important physical values of $n$ are:
\begin{itemize}
\item $n=2$. In this case we can view each loop as being oriented
in either a clockwise or anti-clockwise sense, giving it an overall
weight 2. Each loop configuration then corresponds to a
configuration of integer valued\ \em height \em variables $h(R)$
on the dual lattice, with the convention that the nearest
neighbour difference $h(R')-h(R)$ takes the values 0, $+1$ or $-1$
according to whether the edge crossed by $RR'$ is unoccupied,
occupied by an edge oriented at $90^{\circ}$ to $RR'$, or at
$-90^{\circ}$. (That is, the current running around each loop is
the lattice curl of $h$.) The variables $h(R)$ may be pictured as
the local height of a crystal surface. In the low-temperature
phase (small $x$) the surface is smooth: fluctuations of the
height differences decay exponentially with separation. In the
high-temperature phase it is rough: they grow logarithmically. In
between is a roughening transition. It is believed that relaxing
the above restriction on the height difference $h(R')-h(R)$ does
not change the universality class, as long as large values of this
difference are suppressed, for example using the weighting
$\exp\big[-\beta\big(h(R')-h(R)\big)^2\big]$. This is the discrete
Gaussian model. It is dual to a model of 2-component spins with
O$(2)$ symmetry called the XY model. \item $n=0$. In this case,
closed loops are completely suppressed, and we have a single
non-self-intersecting path connecting $r_1$ and $r_2$, weighted by
its length. Thus, all paths of the same length are counted with
equal weight. This is the self-avoiding walk problem, which is
supposed to describe the behaviour of long flexible polymer
chains. As $x\to x_c-$, the mean length diverges. The region
$x>x_c$ is the dense phase, corresponding to a long polymer whose
length is of the order of the area of the box, so that it has
finite density. \item $n=-2$ corresponds to the loop-erased random
walk. This is an ordinary random walk in which every loop, once it
is formed, is erased. Taking $n=-2$ in the $O(n)$ model of
non-intersecting loops has this effect.
\end{itemize}
\subsection{Potts model.}
Another important model which may described in terms of random
curves in the $Q$-state Potts model. This is most easily considered
on square lattice, at each site of which is a variable $s(r)$ which can take
$Q$ (initially a positive integer) different values. The partition function
is
\begin{equation}
Z_{\rm Potts}={\rm Tr}\,e^{\beta J\sum_{rr'}\delta_{s(r),s(r')}}
\propto{\rm
Tr}\,\prod_{rr'}\left(1-p+p\delta_{s(r),s(r')}\right)\,,
\end{equation}
with $e^{\beta J}=(1-p)^{-1}$. The product may be expanded in a
similar way to the case of the Ising model. All possible graphs
$\cal G$ will appear. Within each connected component of $\cal G$
the Potts spins must be equal, giving rise to a factor $Q$ when
the trace is performed. The result is
\begin{equation}
Z_{\rm Potts}=\sum_{\cal G}\,p^{|{\cal G}|}(1-p)^{|\overline{\cal
G}|} Q^{||{\cal G}||}\,,
\end{equation}
where $|{\cal G}|$ is the number of edges in $\cal G$,
$|\overline{\cal G}|$ the number in its complement, and $||{\cal
G}||$ is the number of connected components of $\cal G$, which are
called Fortuin-Kasteleyn (FK) clusters. This is the \em random
cluster \em representation of the Potts model. When $p$ is small,
the mean cluster size is small. As $p\to p_c$, it diverges, and
for $p>p_c$ there is an infinite cluster which contains a finite
fraction of all the sites in the lattice. It should be noted that
these FK clusters are \em not \em the same as the spin clusters
within which the original Potts spins all take the same value.

The limit $Q\to1$ gives another realisation of percolation -- this
time bond percolation on the square lattice. For $Q\to0$ there is
only one cluster. If at the same time $x\to0$ suitably, all loops
are suppressed and the only graphs $\cal G$ which contribute are
\em spanning trees\em, which contain every site of the lattice. In
the Potts partition function each possible spanning tree is
counted with the same weight, corresponding to the problem of
uniform spanning trees (UST). The ensemble of paths on USTs
connecting two points $r_1$ and $r_2$ turns out to be be that of
loop-erased random walks.
\begin{figure}
\centering
\includegraphics[width=8cm]{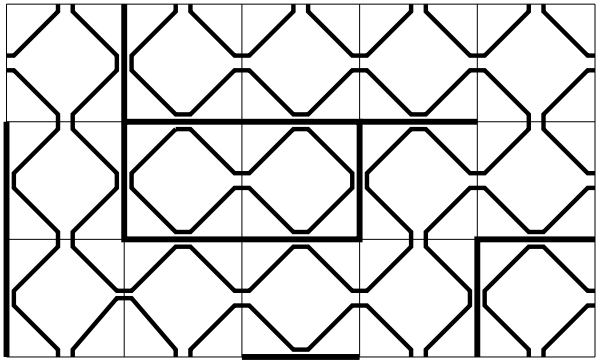}
\caption{\label{fig:medial}\small Example of FK clusters (heavy
lines) in the random cluster representation of the Potts model,
and the corresponding set of dense loops (medium heavy) on the
medial lattice. The loops never cross the edges connecting sites
in the same cluster.}
\end{figure}

The random cluster model may be realised as a gas of dense loops
in the way illustrated in Fig.~\ref{fig:medial}. These loops lie
on the medial lattice, which is also square but has twice the
number of sites. It may be shown that, at $p_c$, the weights for
the clusters are equivalent to counting each loop with a fugacity
$\sqrt Q$. Thus the boundaries of the critical FK clusters in the
$Q$-state Potts model are the same in the scaling limit (if it
exists) as the closed loops of the dense phase of the O$(n)$
model, with $n=\sqrt Q$.

To generate an open path in the random cluster model
connecting sites $r_1$ and $r_2$ on the boundary we must choose
`wired' boundary conditions, in which $p
=1$ on all the edges
parallel to the boundary, from $r_1$ to $r_2$, and free boundary
conditions, with $p=0$, along the remainder.

\subsection{Coulomb gas methods}\label{sec:coulomb}
Many important results concerning the O$(n)$ model can be derived
in a non-rigorous fashion using so-called Coulomb gas methods. For
the purposes of comparison with later results from SLE, we now
summarise these methods and collect a few relevant formulae. A
much more complete discussion may be found in the review by
Nienhuis\cite{Nienhuis}.

We assume that the boundary conditions on the O$(n)$ spins are free,
so that the partition function is a sum over closed loops only.
First orient each loop at random. Rather than giving clockwise and
anti-clockwise orientations the same weight $n/2$, give them complex
weights $e^{\pm 6i\chi}$, where $n=e^{6i\chi}+e^{-6i\chi}
=2\cos6\chi$. These may be taken into account, on the honeycomb lattice,
by assigning a weight $e^{\pm i\chi}$ at each vertex where an oriented
loop turns R (resp.~L). This transforms the non-local factors of $n$
into local (albeit complex) weights depending only on the local
configuration at each vertex.

Next transform to the height variables described above. By convention,
the heights are taken to be integer multiples of $\pi$. The local weights
at each vertex now depend only on the differences of the three adjacent
heights. The crucial assumption of the Coulomb gas approach is that,
under the RG, this model flows to one in which the lattice can be replaced
by a continuum, and the heights go over into a gaussian free field, with
partition function $Z=\int e^{-S[h]}[dh]$, where
\begin{equation}\label{gff}
S=(g/4\pi)\int(\nabla h)^2d^2r\,.
\end{equation}
As it stands, this is a simple free field theory.
The height fluctuations grow logarithmically:
$\langle\big(h(r_1)-h(r_2)\big)^2\rangle\sim(2/g)\ln|r_1-r_2|$,
and the correlators of exponentials of the height decay with power
laws:
\begin{equation}
\langle e^{iqh(r_1)}\,e^{-iqh(r_2)}\rangle\sim
|r_1-r_2|^{-2x_q}\,,
\end{equation}
where $x_q=q^2/2g$. All the subtleties come from the combined
effects of the phase factors and the boundaries or the topology.
This is particularly easy to see if we consider the model on a
cylinder of circumference $\ell$ and length $L\gg\ell$. In the
simple gaussian model (\ref{gff}) the correlation function between
two points a distance $L$ apart along the cylinder decays as
$\exp\big(-2\pi x_qL/\ell\big)$. However, if $\chi\not=0$, loops
which wrap around the cylinder are not counted correctly by the
above prescription, because the total number of left turns minus
right turns is then zero. We may arrange the correct factors by
inserting $e^{\pm 6i\chi h/\pi}$ at either end of the cylinder.
This has the effect of modifying the partition function: one finds
$\ln Z\sim (\pi c/6)(L/\ell)$ with
\begin{equation}\label{eqn:c6}
c=1-6{(6\chi/\pi)^2\over g}\,.
\end{equation}
This dependence of the partition function is one way of
determining the so-called central charge of the corresponding CFT
(Sec.~\ref{sec:cft}). The charges at each end of the cylinder also
modify the scaling dimension $x_q$ to
$(1/2g)\big((q-6i\chi/\pi)^2- (6i\chi/\pi)^2\big)$.

The value of $g$ may be fixed \cite{Kondev} in terms of the
original discreteness of the height variables as follows: adding a
term $-\lambda\int\cos2h\,d^2r$ to $S$ in (\ref{gff}) ensures
that, in the limit $\lambda\to\infty$, $h$ will be an integer
multiple of $\pi$. For this deformation not to affect the critical
behaviour, it must be marginal in the RG sense, which means that
it must have scaling dimension $x_2=2$. This condition then
determines $g=1-6\chi/\pi$.

\subsubsection{Winding angle distribution}\label{sec:winding1}
A simple property which can be inferred from the Coulomb gas
formulation is the winding angle distribution. Consider a cylinder
of circumference $2\pi$ and a path that winds around it. What the
probability that it winds through an angle $\theta$ around the
cylinder while it moves a distance $L\gg1$ along the axis? This
will correspond to a height difference $\Delta h=\pi(\theta/2\pi)$
between the ends of the cylinder, and therefore an additional free
energy $(g/4\pi)(2\pi L)(\theta/2L)^2$. The probability density is
therefore
\begin{equation}
P(\theta)\propto \exp(-g\theta^2/8L)\,,
\end{equation}
so that $\theta$ is normally distributed with variance $(4/g)L$.
This result will be useful later (Sec.~\ref{sec:radialsle}) for
comparison with SLE.

\subsubsection{$N$-leg exponent}\label{Nleg1}
As a final simple exponent prediction, consider the correlation
function $\langle\Phi_N(r_1)\Phi_N(r_2)\rangle$ of the $N$-leg
operator, which in the language of the O$(n)$ model is
$\Phi_N=s_{a_1}\cdots s_{a_N}$, where none of the indices are
equal. It gives the probability that $N$ mutually non-intersecting
curves connect the two points. Taking them a distance $L\gg\ell$
apart along the cylinder, we can choose to orient them all in the
same sense, corresponding to a discontinuity in $h$ of $N\pi$ in
going around the cylinder. Thus we can write $h=\pi Nv/\ell+\tilde
h$, where $0\leq v<\ell$ is the coordinate around the cylinder,
and $\tilde h(v+\ell)=\tilde h(v)$. This gives
\begin{equation}\label{eqn:Nleg}
\langle\Phi_N(r_1)\Phi_N(r_2)\rangle\sim
e^{-(g/4\pi)(N\pi/\ell)^2L+(\pi L/6\ell)-(\pi cL/6\ell)}\,.
\end{equation}
The second term in the exponent comes from the integral over the
fluctuations $\tilde h$, and the last from the partition function.
They differ because in the numerator, once there are curves
spanning the length of the cylinder, loops around it, which give
the correction term in (\ref{eqn:c6}), are forbidden.
(\ref{eqn:Nleg}) then gives
\begin{equation}
x_N=(gN^2/8)-(g-1)^2/2g\,.
\end{equation}
\newpage
\section{SLE}\label{sec:sle}
\subsection{The postulates of SLE.}\label{sec:postulates}
SLE gives a description of the continuum limit of the lattice
curves connecting two points on the boundary of a domain $\cal D$
which were introduced in Sec.~\ref{sec:lattice}. The idea is to
define a \em measure \em $\mu(\gamma;{\cal D},r_1,r_2))$ on these
continuous curves. (Note that the notion of a probability density
of such objects does not make sense, but the more general concept
of a measure does.)

There are two basic properties of this
continuum limit which must either be assumed, or, better, proven to hold
for a particular lattice model.
The first is the continuum version of Property 1:

{\bf Property 3.1} (continuum version). Denote the curve by
$\gamma$, and divide it into two disjoint parts: $\gamma_1$ from
$r_1$ to $\tau$, and $\gamma_2$ from $\tau$ to $r_2$. Then the
conditional measure $\mu(\gamma_2|\gamma_1;{\cal D},r_1,r_2)$ is
the same as $\mu(\gamma_2;{\cal D}\setminus\gamma_1,\tau,r_2)$.

This property we expect to be true for the scaling limit
of all such curves in the
O$(n)$ model (at least for $n\geq0$), even away from the critical
point. However the second property encodes the notion of conformal
invariance, and it should be valid, if at all, only at $x=x_c$
and, separately, for $x>x_c$.

{\bf Property 3.2} (conformal invariance.) Let $\Phi$ be a
conformal mapping of the interior of the domain $\cal D$ onto the
interior of ${\cal D}'$, so that the points $(r_1,r_2)$ on the
boundary of $\cal D$ are mapped to points $(r_1',r_2')$ on the
boundary of ${\cal D}'$. The measure $\mu$ on curves in $\cal D$
induces a measure $\Phi*\mu$ on the image curves in ${\cal D}'$.
The conformal invariance property states that this is the same as
the measure which would be obtained as the continuum limit of
lattice curves from $r_1'$ to $r_2'$ in ${\cal D}'$. That is
\begin{equation}
(\Phi*\mu)(\gamma;{\cal D},r_1,r_2)= \mu(\Phi(\gamma);{\cal
D}',r_1',r_2')\,.
\end{equation}
\subsection{Loewner's equation}\label{sec:loewner}
We have seen that, on the lattice, the curves $\gamma$ may be
`grown' through a discrete exploration process. The Loewner
process is the continuum version of this. Because of Property 2 it
suffices to describe this in a standard domain $\cal D$, which is
taken to be the upper half plane $\bf H$, with the points $r_1$
and $r_2$ being the origin and infinity respectively.

\begin{figure}
\centering
\includegraphics[width=6cm]{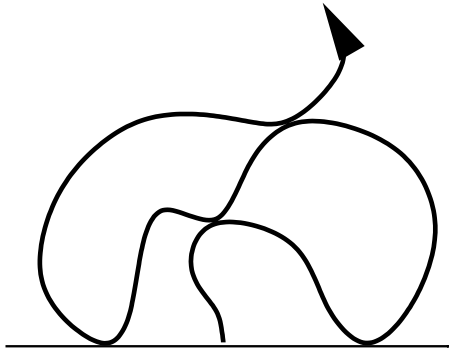}
\caption{\label{fig:hull}\small Schematic view of a trace and its
hull.}
\end{figure}

The first thing to notice is that, although on the honeycomb
lattice the growing path does not intersect itself, in the
continuum limit it might (although it still should not cross
itself.) This means that there may be regions enclosed by the path
which are not on the path but nevertheless are not reachable from
infinity without crossing it. We call the union of the set of such
points, together with the curve itself, up to time $t$, the \em
hull \em $K_t$. (This is a slightly different usage of this term
from that in the physics percolation literature.) It is the
complement of the connected component of the half plane which
includes $\infty$, itself denoted by ${\bf H}\setminus K_t$. See
Fig.~\ref{fig:hull}.

Another property which often holds in the half-plane is that of
\em reflection invariance\em: the distribution of lattice paths
starting from the origin and ending at $\infty$ is invariant under
$x\to-x$. For the lattice paths in the $O(n)$ model discussed in
Sec.~\ref{sec:Onmodel} this follows from the symmetry of the
underlying weights, but for the boundaries of the FK clusters in
the Potts model it is a consequence of duality. Not all simple
curves in lattice models have this property. For example if we
consider the 3-state Potts model in which the spins on the
negative and positive real axes are fixed to different values,
there is a simple lattice curve which forms the outer boundary of
the spin cluster containing the positive real axis. This is not
the same as the boundary of the spin cluster containing the
negative real axis, and it is not in general symmetric under
reflections.

Since ${\bf H}\setminus K_t$ is simply connected, by the Riemann
mapping theorem it can be mapped into the standard domain $\bf H$
by an analytic function $g_t(z)$. Because this preserves the real
axis outside $K_t$ it is in fact real analytic. It is not unique,
but can be made so by imposing the behaviour as $z\to\infty$
\begin{equation}\label{eqn:asymp}
g_t(z)\sim z+O(1/z)\,.
\end{equation}
It can be shown that, as the path grows, the coefficient of $1/z$
is monotonic increasing (essentially it is the electric dipole
moment of $K_t$ and its mirror image in the real axis.) Therefore
we may reparametrise time so that this coefficient is $2t$. (The
factor 2 is conventional.) Note that the length of the curve is
not be a useful parametrisation in the continuum limit, since the
curve is a fractal.

The function $g_t(z)$ maps the whole boundary of $K_t$ onto part
of the real axis. In particular, it maps the growing tip $\tau_t$
to a real point $a_t$. Any point on the real axis outside $K_t$
remains on the real axis. As the path grows, the point $a_t$ moves
on the real axis. For the path to describe a curve, it must always
grow only at its tip, and this means that the function $a_t$ must
be continuous, but not necessarily differentiable.

A simple but instructive example is when $\gamma$ is a straight
line growing vertically upwards from a fixed point $a$. In this
case
\begin{equation}\label{eqn:example}
g_t(z)=a+\big((z-a)^2+4t\big)^{1/2}\,.
\end{equation}
This satisfies (\ref{eqn:asymp}), and $\tau_t=2i\sqrt t$. More
complicated deterministic examples can be found\cite{KadKager}. In
particular, $a_t\propto t^{1/2}$ describes a straight line growing
at a fixed angle to the real axis.

Loewner's idea \cite{Loewner} was to describe the path $\gamma$
and the evolution of the tip $\tau_t$ in terms of the evolution of
the conformal mapping $g_t(z)$. It turns out that the equation of
motion for $g_t(z)$ is simple:
\begin{equation}\label{eqn:loewner}
{dg_t(z)\over dt}={2\over g_t(z)-a_t}\,.
\end{equation}
This is Loewner's equation. The idea of the proof is
straightforward. Imagine evolving the path for a time $t$, and
then for a further short time $\delta t$. The image of
$K_{t+\delta t}$ under $g_t$ is a short vertical line above the
point $a_t$ on the real axis. Thus we can write, using
(\ref{eqn:example})
\begin{equation}
g_{t+\delta t}(z)\approx a_t+\big((g_t(z)-a_t)^2+4\delta
t\big)^{1/2}\,.
\end{equation}
Differentiating with respect to $\delta t$ and then letting
$\delta t\to0$, we obtain (\ref{eqn:loewner}).

Note that, even if $a_t$ is not differentiable (as is the case for
SLE), (\ref{eqn:loewner}) gives for each point $z_0$ a solution
$g_t(z_0)$ which is differentiable with respect to $t$, up to the
time when $g_t(z_0)=a_t$. This is the time when $z_0$ is first
included in $K_t$. However, it is sometimes (see
Sec.~\ref{sec:cft}) useful to normalise the Loewner mapping
differently, defining $\hat g_t(z)=g_t(z)-a_t$, which always maps
the growing tip $\tau_t$ to the origin. If $a_t$ is not
differentiable, neither is $\hat g_t$, and the Loewner equation
should be written in differential form as $d\hat g_t=(2dt/\hat
g_t)-da_t$.

Given a growing path, we can determine the hull $K_t$ and hence,
in principle, the function $g_t(z)$ and thereby $a_t=g_t(\tau_t)$.
Conversely, given $a_t$ we can integrate (\ref{eqn:loewner}) to
find $g_t(z)$ and hence in determine the curve (although proving
that this inverse problem gives a curve is not easy.)

\subsection{Schramm-Loewner Evolution}\label{sec:schramm}
In the case that we are interested in, $\gamma$ is a random curve,
so that $a_t$ is a random continuous function. What is the measure
on $a_t$? This is answered by the following remarkable result, due
to Schramm\cite{Schramm}:

{\bf Theorem.} \em If Properties 3.1 and 3.2 hold, together with
reflection symmetry, then $a_t$ is proportional to a standard
Brownian motion.\em

That is
\begin{equation}
a_t=\sqrt\kappa B_t\,,
\end{equation}
so that $\langle a_t\rangle=0$,
$\langle\big(a_{t_1}-a_{t_2}\big)^2\rangle=\kappa|t_1-t_2|$. The
only undetermined parameter is $\kappa$, the diffusion constant.
It will turn out that different values of $\kappa$ correspond to
different universality classes of critical behaviour.

\begin{figure}
\centering
\includegraphics[width=8cm]{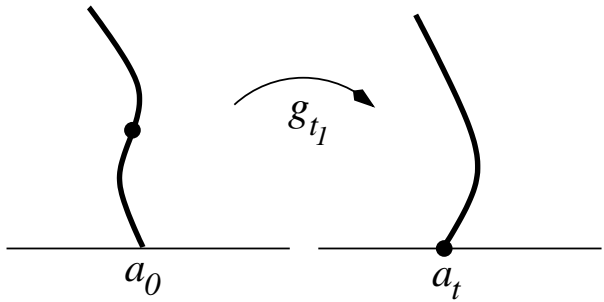}
\caption{\label{fig:loewner}\small A hull evolved from $a_0$ for
time $t_1$, then to infinity. The measure on the image of the rest
of the curve under $g_{t_1}$ is the same, according to the
postulates of SLE, as a hull evolved from $a_{t_1}$ to $\infty$.}
\end{figure}

The idea behind the proof is once again simple. As before,
consider growing the curve for a time $t_1$, giving $\gamma_1$,
and denote the remainder $\gamma\setminus\gamma_1=\gamma_2$.
Property 3.1 tells us that the conditional measure on $\gamma_2$
given $\gamma_1$ is the same as the measure on $\gamma_2$ in the
domain ${\bf H}\setminus K_{t_1}$, which, by Property 3.2, induces
the same measure on $g_{t_1}(\gamma_2)$ in the domain ${\bf H}$,
shifted by $a_{t_1}$ (see Fig.~\ref{fig:loewner}).
In terms of the function $a_t$ this means that the probability law
of $a_t-a_{t_1}$, for $t>t_1$, is the same as the law of
$a_{t-t_1}$. This implies that all the increments $\Delta_n \equiv
a_{(n+1)\delta t}-a_{n\delta t}$ are independent identically
distributed random variables, for all $\delta t>0$. The only
process that satisfies this is Brownian motion with a possible
drift term: $a_t=\sqrt\kappa B_t+\alpha t$. Reflection symmetry
then implies that $\alpha=0$.

\subsection{Simple Properties of SLE}\label{sec:simpleproperties}
\subsubsection{Phases of SLE}\label{sec:phases}
Many of the results discussed in this section have been proved by
Rohde and Schramm\cite{RS}. First we address the question of how
the trace (the trajectory of $\tau_t$) looks for different values
of $\kappa$. For $\kappa=0$, it is a vertical straight line. As
$\kappa$ increases, the trace should randomly turn to the L or R
more frequently. However, it turns out that there are qualitative
differences at critical values of $\kappa$. To see this, let us
first study the process on the real axis. Let
$x_t=g_t\big(x_0\big)-a_t$ be the distance between the image at
time $t$ of a point which starts at $x_0$ and the image $a_t$ of
the tip. It obeys the stochastic equation
\begin{equation}\label{eqn:bessel}
dx_t={2dt\over x_t}-\sqrt\kappa dB_t\,.
\end{equation}
Physicists often write such an equation as $\dot x=(2/x)-\eta_t$
where $\eta_t$ is `white noise' of strength $\kappa$. Of course
this does not make sense since $x_t$ is not differentiable. Such
equations are always to be interpreted in the `Ito sense', that
is, as the limit as $\delta t\to0$ of the forward difference
equation $x_{t+\delta t}\approx x_t+(2\delta
t/x_t)+\int_t^{t+\delta t}\eta_{t'}dt'$.

(\ref{eqn:bessel}) is known as the Bessel process. (If we set
$R_t=(D-1)^{1/2}x_t/2$ and $\kappa^2=4/(D-1)$ it describes the
distance $R_t$ from the origin of a Brownian particle in $D$
dimensions.) The point $x_t$ is repelled from the origin but it is
also subject to a random force. Its ultimate fate can be inferred
from the following crude argument: if we ignore the random force,
$x_t^2\sim 4t$, while, in the absence of the repulsive term,
$\langle x_t^2\rangle\sim\kappa t$. Thus for $\kappa<4$ the
repulsive force wins and the particle escapes to infinity, while
for $\kappa>4$ the noise dominates and the particle collides with
the origin in finite time (at which point the equation breaks
down.) A more careful analysis confirms this.
\begin{figure}
\centering
\includegraphics[width=5cm]{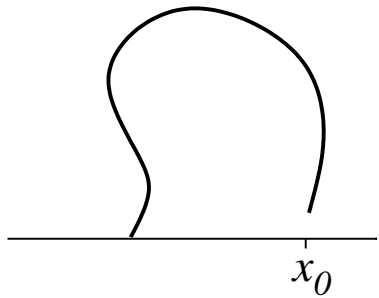}
\caption{\label{fig:enclose}\small The trace is about to hit the
axis at $x_0$ and enclose a region. At the time this happens, the
whole region including the point $x_0$ is mapped by $g_t$ to the
same point $a_t$.}
\end{figure}
What does this collision signify in  terms of the behaviour of the
trace? In Fig.~\ref{fig:enclose} we show a trace which is about to
hit the real axis at the point $x_0$, thus engulfing a whole
region. This is visible from infinity only through a very small
opening, which means that, under $g_t$, it gets mapped to a very
small region. In fact, as the tip $\tau_t$ approaches $x_0$, the
size of the image of this region shrinks to zero. When the gap
closes, the whole region enclosed by the trace, as well as
$\tau_t$ and $x_0$, are mapped in to the single point $a_t$, which
means, in particular, that $x_t\to0$. The above argument shows
that for $\kappa<4$ this never happens: the trace never hits the
real axis (with probability 1.) For the same reason, it neither
hits itself. Thus for $\kappa<4$ the trace $\gamma$ is a \em
simple curve\em.

The opposite is true for $\kappa>4$: points on the real axis are
continually colliding with the image $a_t$ of the tip. This means
that the trace is continually touching both itself and the real
axis, at the same time engulfing whole regions. Moreover, since it
is self-similar object, it does this on all scales, an infinite
number of times within any finite neighbourhood! Eventually the
trace swallows the whole half plane: every point is ultimately
mapped into $a_t$. For $\kappa<4$ only the points on the trace
itself suffer this fate. The case $\kappa=4$ is more tricky: in
fact the trace is then also a simple curve.

When $\kappa$ is just above 4, the images of points on the real
axis under $g_t$ collide with $a_t$ only when there happen to be
rare events when the random force is strong enough to overcome the
repulsion. When this happens, whole segments of the real axis are
swallowed at one time, corresponding to the picture described
above. Conversely, for large $\kappa$, the repulsive force is
negligible except for very small $x_t$. In that case, two
different starting points move with synchronised Brownian motions
until the one which started off closer to the origin is swallowed.
Thus the real line is eaten up in a continuous fashion rather than
piecemeal. There are no finite regions swallowed by the trace
which are not on the trace itself. This means that the trace is
\em space-filling\em: $\gamma$ intersects every neighbourhood of
every point in the upper half plane. We shall argue later
(Sec.~\ref{sec:fractaldim}) that the fractal dimension of the
trace is $d_f=1+\kappa/8$ for $\kappa\leq8$ and 2 for
$\kappa\geq8$. Thus it becomes space-filling for all
$\kappa\geq8$.

\subsubsection{SLE duality}\label{sec:duality}
For $\kappa>4$ the curve continually touches itself and therefore
its hull $K_t$ contains earlier portions of the trace (see
Fig.~\ref{fig:hull}). However, the \em frontier \em of $K_t$ (i.e.
the boundary of ${\bf H}\setminus K_t$, minus any portions of the
real axis), is by definition a simple curve. A beautiful result,
first suggested by Duplantier\cite{Dupduality}, and proved by
Beffara\cite{Beffaraduality} for the case $\kappa=6$, is that
locally this curve is an SLE$_{\tilde\kappa}$, with
\begin{equation}
\tilde\kappa=16/\kappa\,.
\end{equation}
For example, the exterior of a percolation cluster contains many
`fjords' which, on the lattice, are connected to the main ocean by
a neck of water which is only a finite number of lattice spacings
wide. These are sufficiently frequent and the fjords
macroscopically large that they survive in the continuum limit.
SLE$_{6}$ describes the boundaries of the clusters, including the
coastline of all the fjords. However, the coastline as seen from
the ocean is a simple curve, which is locally SLE$_{8/3}$, the
same as that conjectured for a self-avoiding walk. This suggests,
for example, that locally the frontier of a percolation cluster
and a self-avoiding walk are the same in the scaling limit. In
Sec.~\ref{sec:cft} we show that SLE$_\kappa$ and
SLE$_{\tilde\kappa}$ correspond to CFTs with the same value of the
central charge $c$.

\subsection{Special values of $\kappa$}
\subsubsection{Locality}\label{sec:locality}
\begin{figure}
\centering
\includegraphics[width=10cm]{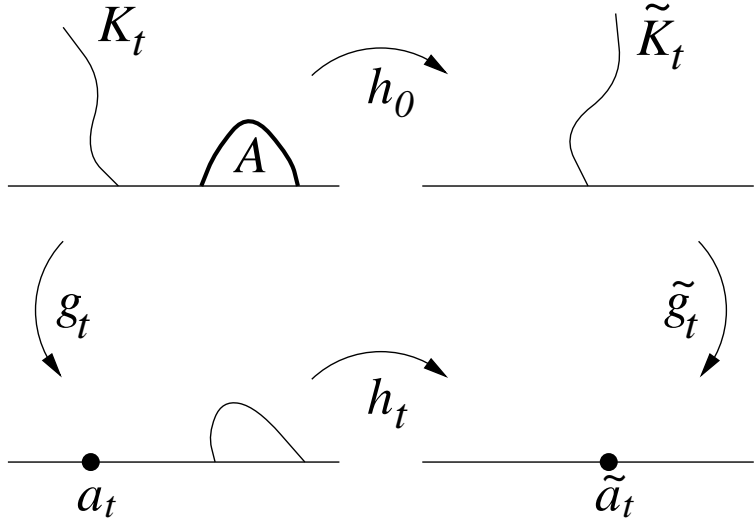}
\caption{\label{fig:bump}\small An SLE hull in ${\bf H}\setminus
A$ and two different ways of removing it: either by first removing
$A$ through $h_0$ and then using a Loewner map $\tilde g_t$ in the
image of ${\bf H}\setminus A$; or by removing $K_t$ first with
$g_t$ and then removing the image of $A$ with $h_t$. Since all
maps are normalised, this diagram commutes.}
\end{figure}

[This subsection and the next are more technical and may be
omitted at a first reading.] We have defined SLE in terms of
curves which connect the origin and infinity in the upper half
plane. Property 2 then allows us to define it for any pair of
boundary points in any simply connected domain, by a conformal
mapping. It is interesting to study how the variation of the
domain affects the SLE equation. Let $A$ be a simply connected
region connected to the real axis which is at some finite distance
from the origin (see Fig.~\ref{fig:bump}).
Consider a trace $\gamma_t$, with hull $K_t$, which grows from the
origin according to SLE in the domain ${\bf H}\setminus A$.
According to Property 2, we can do this by first making a
conformal mapping $h_0$ which removes $A$, and then a map $\tilde
g_t$ which removes the image $\tilde K_t=h_0(K_t)$. This would be
described by SLE in $h_0({\bf H}\setminus A)$, except that the
Loewner `time' would not in general be the same as $t$. However,
another way to think about this is to first use a SLE map $g_t$ in
$\bf H$ to remove $K_t$, then another map, call it $h_t$, to
remove $g_t(A)$. Since both these procedures end up removing
$K_t\cup A$, and all the maps are assumed to be normalised at
infinity in the standard way (\ref{eqn:asymp}), they must be
identical, that is $h_t\circ g_t=\tilde g_t\circ h_0$ (see
Fig.~\ref{fig:bump}). If $g_t$ maps the growing tip $\tau_t$ to
$a_t$, then after both mappings it goes to $\tilde a_t=h_t(a_t)$.
We would like to understand the law of $\tilde a_t$.

Rather than working this out in full generality (see for example
\cite{Wreview}), let us suppose that $A$ is a short vertical
segment $(x,x+i\epsilon)$ with $\epsilon\ll x$, and that $t=dt$ is
infinitesimal. Then, under $g_{dt}$, $x\to x+2dt/x$ and
$\epsilon\to\epsilon(1-2dt/x^2)$. The map that removes this is
(see (\ref{eqn:example}))
\begin{equation}\label{eqn:remove}
h_{dt}(z)=\left((z-x-2dt/x)^2+\epsilon^2(1-2dt/x^2)^2\right)^{1/2}
+x+2dt/x\,.
\end{equation}
In order to find $\tilde a_{dt}$, we need to set
$z=a_{dt}=\sqrt\kappa dB_t$ in this expression. Carefully
expanding this to first order in $dt$, remembering that
$(dB_t)^2=dt$, and also taking the first non-zero contribution in
$\epsilon/x$, gives after a few lines of algebra
\begin{equation}\label{eqn:kappa6}
\tilde a_{dt}=(1-\epsilon^2/x^2)\sqrt\kappa
dB_t+\ffrac12(\kappa-6)(\epsilon^2/x^3)dt\,.
\end{equation}
The factor in front of the stochastic term may be removed by
rescaling $dt$: this restores the correct Loewner time. But there
is also a drift term, corresponding to the effect of $A$. For
$\kappa<6$ we see that the SLE is initially repelled from $A$.
From the point of view of the exploration process for the Ising
model discussed in Sec.~\ref{sec:exploration}, this makes sense:
if the spins along the positive real axis and on $A$ are fixed to
be up, then the spin just above the origin is more likely to be up
than down, and so $\gamma$ is more likely to turn to the left.

For $\kappa=6$, however, this is no longer the case: the presence
of $A$ does not affect the initial behaviour of the curve. This is
a particular case of the property of \em locality \em when
$\kappa=6$, which states that, for any $A$ as defined above, the
law of $K_t$ in ${\bf H}\setminus A$ is, up to a time
reparametrisation, the same as the law of $K_t$ in $\bf H$, as
long as $K_t\cap A=\emptyset$. That is, up to the time that the
curve hits $A$, it doesn't know it's there. Such a property would
be expected for the cluster boundaries of uncorrelated Ising spins
on the lattice, i.e.~percolation. This is then consistent with the
identification of percolation cluster boundaries with SLE$_6$.
\subsubsection{Restriction}\label{sec:restriction}
It is also interesting to work out how the local scale transforms
in going from $a_t$ to $\tilde a_t$. A measure of this is
$h'_t(a_t)$. A similar calculation starting from
(\ref{eqn:remove}) gives, in the same limit as above,
\begin{equation}
d\big(h'(a_t)\big)=
h'_{dt}(a_{dt})-h_0'(0)=(\epsilon^2/x^3)\sqrt\kappa
dB_t+\ffrac12\left(
(\epsilon^4/x^6)+(\kappa-\ffrac83)(3\epsilon^2/x^4)\right)\,.
\end{equation}
Now something special happens when $\kappa=\frac83$. The drift
term in $d\big(h'(a_t)\big)$ does not then vanish, but if we take
the appropriate power $d\big(h'_t(a_t)^{5/8}\big)$ it does. This
implies that the \em mean \em of $h'_t(a_t)^{5/8}$ is conserved.
Now at $t=0$ it takes the value $\Phi_A'(0)^{5/8}$, where
$\Phi_A=h_0$ is the map that removes $A$. If $K_t$ hits $A$ at
time $T$ it can be seen from (\ref{eqn:remove}) that $\lim_{t\to
T}h'_t(a_t)^{5/8}=0$. On the other hand, if it never hits $A$ then
$\lim_{t\to\infty}h'_t(a_t)^{5/8}=1$. Therefore $\Phi_A'(0)^{5/8}$
gives the \em probability that the curve $\gamma$ does not
intersect $A$\em.

This is a remarkable result in that it depends only on the value
of $\Phi'_A$ at the starting point of the SLE (assuming of course
that $\Phi_A$ is correctly normalised at infinity.) However it has
the following even more interesting consequence. Let
$\hat\Phi_A(z)=\Phi_A(z)-\Phi(0)$. Consider the ensemble of all
SLE$_{8/3}$ in $\bf H$, and the sub-ensemble consisting of all
those curves $\gamma$ which do not hit $A$. Then the measure on
the image $\hat\Phi_A(\gamma)$ in $\bf H$ is again given by
SLE$_{8/3}$. The way to show this is to realise that the measure
on $\gamma$ is characterised by the probability $P(\gamma\cap
A'=\emptyset)$ that $\gamma$ does not hit $A'$ for all possible
$A'$. The probability that $\hat\Phi_A(\gamma)$ does not hit $A'$,
given that $\gamma$ does not hit $A$, is the ratio of the
probabilities $P(\gamma\cap{\hat\Phi_A}^{-1}(A')=\emptyset)$ and
$P(\gamma\cap A=\emptyset)$. By the above result, the first factor
is the derivative at the origin of the map
$\hat\Phi_{A'}\circ\hat\Phi_A$ which removes $A$ then $A'$, while
the second is the derivative of the map which removes $A$. Thus
\begin{equation}
P(\hat\Phi_A(\gamma)\cap A'=\emptyset|\gamma\cap A=\emptyset)=
\left({(\hat\Phi_{A'}\circ\hat\Phi_A)'(0)\over
{\hat\Phi_A}'(0)}\right)^{5/8}={\hat\Phi_{A'}}'(0)^{5/8}
=P(\gamma\cap A'=\emptyset)\,.
\end{equation}
Since this is true for all $A'$, it follows that the law of
$\hat\Phi_A(\gamma)$ given that $\gamma$ does not intersect $A$ is
the same as that of $\gamma$. This is called the \em restriction
property\em. Note that while, according to Property 2, the law of
an SLE in any simply connected subset of $\bf H$ is determined by
the conformal mapping of this subset to $\bf H$, the restriction
property is stronger than this, and it holds only when
$\kappa=\frac83$.

We expect such a property to hold for the continuum limit of
self-avoiding walks, assuming it exists. On the lattice, every
walk of the same length is counted with the same weight. That is,
the measure is uniform. If we consider the sub-ensemble of such
walks which avoid a region $A$, the measure on the remainder
should still be uniform. This will be true if the restriction
property holds. This supports the identification of self-avoiding
walks with SLE$_{8/3}$.

\subsection{Radial SLE and the winding angle}\label{sec:radialsle}
So far we have discussed a version of SLE that gives a conformally
invariant measure on curves which connect two distinct boundary
points of a simply connected domain $\cal D$. For this reason it
is called \em chordal \em SLE. There are variants which describe
other situations. For example, one could consider curves $\gamma$
which connect a given point $r_1$ on the boundary to an interior
point $r_2$. The Riemann mapping theorem allows us to map
conformally onto another simple connected domain, with $r_2$ being
mapped to any preassigned interior point. It is simplest to choose
for the standard domain the unit disc $\bf U$, with $r_2$ being
mapped to the origin. So we are considering curves $\gamma$ which
connect a given point $e^{i\theta_0}$ on the boundary with the
origin. As before, we may consider growing the curve dynamically.
Let $K_t$ be the hull of that portion which exists up to time $t$.
Then there exists a conformal mapping $g_t$ which takes ${\bf
U}\setminus K_t$ to $\bf U$, such that $g_t(0)=0$. There is one
more free parameter, which corresponds to a global rotation: we
use this to impose the condition that $g_t'(0)$ is real and
positive. One can then show that, as the curve grows, this
quantity is monotonically increasing, and we can use this to
reparametrise time so that $g_t'(0)=e^t$. This normalised mapping
then takes the growing tip $\tau_t$ to a point $e^{i\theta_t}$ on
the boundary.

Loewner's theorem tells us that $\dot g_t(z)/g_t(z)$, when
expressed as a function of $g_t(z)$, should be holomorphic in
$\overline{{\bf U}}$ apart from a simple pole at $e^{i\theta_t}$.
Since $g_t$ preserves the unit circle outside $K_t$, $\dot
g_t(z)/g_t(z)$ should be pure imaginary when $|g_t(z)|=1$, and in
order that $g_t'(0)=e^t$, it should approach 1 as $g_t(z)\to0$.
The only possibility is
\begin{equation}\label{eqn:radialsle}
{dg_t(z)\over dt}=-g_t(z)\,{g_t(z)+e^{i\theta_t} \over
g_t(z)-e^{i\theta_t}}\,.
\end{equation}
This is the radial Loewner equation. In fact this is the version
considered by Loewner\cite{Loewner}.

It can now be argued, as before, that given Properties 1 \& 2
(suitably reworded to cover the case when $r_2$ is an interior
point) together with reflection, $\theta_t$ must be proportional
to a standard Brownian motion. This defines radial SLE. It is not
immediately obvious how the radial and chordal versions are
related. However, it can be shown that, if the trace of radial SLE
hits the boundary of the unit disc at $e^{i\theta_{t_1}}$ at time
$t_1$, then the law of $K_t$ in radial SLE, for $t<t_1$, is the
same chordal SLE conditioned to begin at $e^{i\theta(0)}$ and end
at $e^{i\theta_{t_1}}$, up to a reparametrisation of time. This
assures us that, in using the chordal and radial versions with the
same $\kappa$, we are describing the same physical problem.

However, one feature that the trace of radial SLE possesses which
chordal SLE does not is the property that it can wind around the
origin. The winding angle at time $t$ is simply
$\theta_t-\theta_0$. Therefore it is normally distributed with
variance $\kappa t$. At this point we can make a connection to the
Coulomb gas analysis of the O$(n)$ model in
Sec.~\ref{sec:winding1}, where it was shown that the variance in
the winding angle on a cylinder of length $L$ is asymptotically
$(4/g)L$. A semi-infinite cylinder, parametrised by $w$, is
conformally equivalent to the unit disc by the mapping $z=e^{-w}$.
Asymptotically, ${\rm Re}\,w\to{\rm Re}\,w-t$ under Loewner
evolution. Thus we can identify $L\sim t$ and hence
\begin{equation}\label{eqn:kappag}
\kappa=4/g\,.
\end{equation}
\subsubsection{Identification with lattice
models}\label{sec:identification} This result allows use to make a
tentative identification with the various realisations of the
O$(n)$ model described in Sec.~\ref{sec:Onmodel}. We have, using
(\ref{eqn:kappag}), $n=-2\cos(4\pi/\kappa)$ with
$2\leq\kappa\leq4$ describing the critical point at $x_c$, and
$4<\kappa\leq8$ corresponding to the dense phase. Some important
special cases are therefore:
\begin{itemize}
\item $\kappa=-2$: loop-erased random walks (proven in
\cite{LSWUST}); \item $\kappa=\frac83$: self-avoiding walks, as
already suggested by the restriction property,
Sec.~\ref{sec:restriction}; unproven, but see \cite{LSWSAW} for
many consequences; \item $\kappa=3$: cluster boundaries in the
Ising model, however as yet unproven; \item $\kappa=4$: BCSOS
model of roughening transition (equivalent to the 4-state Potts
model and the double dimer model), as yet unproven; also certain
level lines of a gaussian random field and the `harmonic explorer'
(proven in \cite{SSHE}); also believed to be dual to the
Kosterlitz-Thouless transition in the XY model; \item $\kappa=6$:
cluster boundaries in percolation (proven in \cite{Smirnov});
\item $\kappa=8$: dense phase of self-avoiding walks; boundaries
of uniform spanning trees (proven in \cite{LSWUST}).
\end{itemize}

It should be noted that no lattice candidates for $\kappa>8$, or
for the dual values $\kappa<2$, have been proposed. This possibly
has to do with the fact that, for $\kappa>8$, the SLE trace is not
reversible: the law on curves from $r_1$ to $r_2$ is not the same
as the law obtained by interchanging the points. Evidently, curves
in equilibrium lattice models should satisfy reversibility.
\newpage
\section{Calculating with SLE}\label{sec:calc}
SLE shows that the measure on the continuum limit of single curves
in various lattice models is given in terms of 1d Brownian motion.
However, it is not at all clear how thereby to deduce interesting
physical consequences.  We first describe two relatively simple
computations in two-dimensional percolation which can be done
using SLE.
\subsection{Schramm's formula}\label{sec:schrammformula}
Given a curve $\gamma$ connecting two points $r_1$ and $r_2$ on
the boundary of a domain $\cal D$, what is the probability that it
passes to the left of a given interior point? This is not a
question which is natural in conventional approaches to critical
behaviour, but which is very simply answered within
SLE\cite{Schrammformula}.

As usual, we can consider $\cal D$ to be the upper half plane, and
take $r_1=a_0$ and $r_2$ to be at infinity. The curve is then
described by chordal SLE starting at $a_0$. Label the interior
point by the complex number $\zeta$.

Denote the probability that $\gamma$ passes to the left of $\zeta$
by $P(\zeta,\bar\zeta;a_0)$ (we include the dependence on
$\bar\zeta$ to emphasise the fact that this is a not a holomorphic
function.) Consider evolving the SLE for an infinitesimal time
$dt$. The function $g_{dt}$ will map the remainder of $\gamma$
into its image $\gamma'$, which, however, by Properties 1 \& 2,
will have the same measure as SLE started from
$a_{dt}=a_0+\sqrt\kappa dB_t$. At the same time, $\zeta\to
g_{dt}(\zeta)=\zeta+2dt/(\zeta-a_0)$. Moreover, $\gamma'$ lies to
the left of $\zeta'$ iff $\gamma$ lies to the left of $\zeta$.
Therefore
\begin{equation}
P\big(\zeta,\bar\zeta;a_0\big)= \langle
P\big(\zeta+2dt/(\zeta-a_0),\bar\zeta+2dt/(\bar\zeta-a_0),a_0
+\sqrt\kappa dB_t\big)\rangle\,,
\end{equation}
where the average $\langle\ldots\rangle$\ is over all realisations
of Brownian motion $dB_t$ up to time $dt$. Taylor expanding, using
$\langle dB_t\rangle=0$ and $\langle(dB_t)^2\rangle=dt$, and
equating the coefficient of $dt$ to zero gives
\begin{equation}\label{eqn:pde1}
\left({2\over\zeta-a_0}{\partial\over\partial\zeta}
+{2\over\bar\zeta-a_0}{\partial\over\partial\bar\zeta}
+\frac\kappa 2{\partial^2\over\partial a_0^2}\right)
P\big(\zeta,\bar\zeta;a_0\big)=0\,.
\end{equation}
Thus $P$ satisfies a linear second-order partial differential equation,
typical of conditional probabilities in stochastic differential equations.

By scale invariance $P$ in fact depends only on the angle $\theta$
subtended between $\zeta-a_0$ and the real axis. Thus
(\ref{eqn:pde1}) reduces to an ordinary second-order linear
differential equation, which is in fact hypergeometric. The
boundary conditions are that $P=0$ and $1$ when $\theta=\pi$ and
$0$ respectively, which gives (specialising to $\kappa=6$)
\begin{equation}\label{eqn:schrammformula}
P=\frac12+{\Gamma(\ffrac23)\over\sqrt\pi\Gamma(\ffrac16)}
(\cot\theta) {}_2F_1(\ffrac12,\ffrac23,\ffrac32;-\cot^2\theta)\,.
\end{equation}
Note that this may also be written in terms of a single quadrature
since one solution of (\ref{eqn:pde1}) is $P=$ const.
\subsection{Crossing probability}
Given a critical percolation problem inside a simply connected
domain $\cal D$, what is the probability that a cluster connects
two disjoint segments $AB$ and $CD$ of the boundary? This problem
was conjectured to be conformally invariant and (probably) first
studied numerically in \cite{Langlandsetal}. A formula based on
CFT as well as a certain amount of guesswork was conjectured in
\cite{JCcrossing}. It was proved, for the continuum limit of site
percolation on the triangular lattice, by Smirnov\cite{Smirnov}.

\begin{figure}
\centering
\includegraphics[width=8cm]{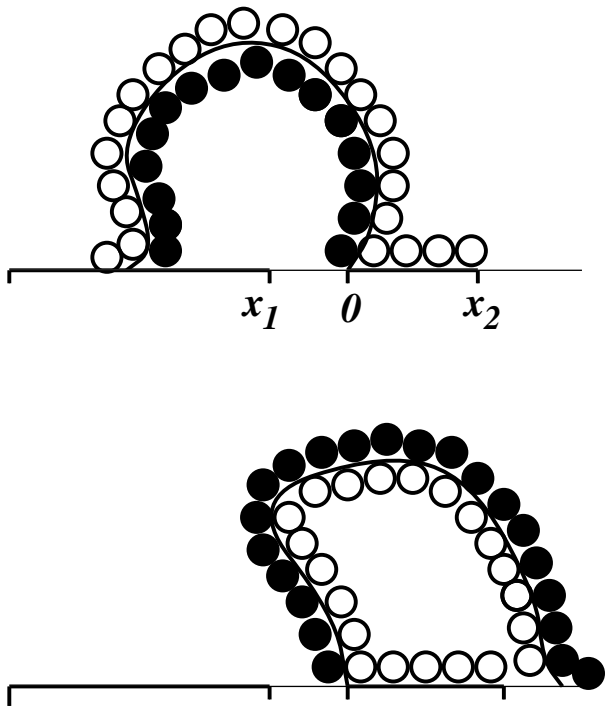}
\caption{\label{fig:crossing}\small Is there a crossing on the
white discs from $(0,x_2)$ to $(-\infty,x_1)$? This happens if and
only if $x_1$ gets swallowed by the SLE before $x_2$.}
\end{figure}

Within SLE, it takes a certain amount of ingenuity\cite{Schramm}
to relate this problem to a question about a single curve. As
usual, let $\cal D$ be the upper half plane. It is always possible
to make a fractional linear conformal mapping which takes $AB$
into $(-\infty,x_1)$ and $CD$ into $(0,x_2)$, where $x_1<0$ and
$x_2>0$. Now go back to the lattice picture and consider critical
site percolation on the triangular lattice in the upper half
plane, so that each site is independently coloured black or white
with equal probabilities $\frac12$. Choose all the boundary sites
on the positive real axis to be white, all those on the negative
real axis to be black (see Fig.~\ref{fig:crossing}). There is a
cluster boundary starting at the origin, which, in the continuum
limit, will be described by SLE$_6$. Since $\kappa>4$, it
repeatedly hits the real axis, both to the L and R of the origin.
Eventually every point on the real axis is swallowed. Either $x_1$
is swallowed before $x_2$, or vice versa.

Note that every site on the L of the curve is black, and every
site on its R is white. Suppose that $x_1$ is swallowed before
$x_2$. Then, at the moment it is swallowed, there exists a
continuous path on the white sites, just to the R of the curve,
which connects $(0,x_2)$ to the row just above $(-\infty,x_1)$. On
the other hand, if $x_2$ is swallowed before $x_1$, there exists a
continuous path on the black sites, just to the L of the curve,
connecting $0-$ to a point on the real axis to the R of $x_2$.
This path forms a barrier (as in the game of Hex) to the
possibility of a white crossing from $(0,x_2)$ to $(-\infty,x_1)$.
Hence there is such a crossing if and only if $x_1$ is swallowed
before $x_2$ by the SLE curve.

Recall that in Sec.~\ref{sec:phases} we related the swallowing of
a point $x_0$ on the real axis to the vanishing of
$x_t=g_t(x_t)-a_t$, which undergoes a Bessel process on the real
line. Therefore
\begin{equation}
{\rm Pr}\big({\rm crossing\ from\ }(0,x_2)\ {\rm to\
}(-\infty,x_1)\big) ={\rm Pr}\big(x_{1t}\ {\rm  vanishes\ before\
} x_{2t}\big)\,.
\end{equation}

Denote this by $P\big(x_1,x_2\big)$. By generalising the SLE to
start at $a_0$ rather than $0$, we can write a differential
equation for this in similar manner to (\ref{eqn:pde1}):
\begin{equation}\label{eqn:pde2}
\left({2\over x_1-a_0}{\partial\over\partial x_1} +{2\over
x_2-a_0}{\partial\over\partial x_2} +\frac\kappa
2{\partial^2\over\partial a_0^2}\right)\
P\big(x_1,x_2;a_0\big)=0\,.
\end{equation}
Translational invariance implies that we can replace
$\partial_{a_0}$ by $-(\partial_{x_1}+\partial_{x_2})$. Finally,
$P$ can in fact depend only on the ratio
$\eta=(x_2-a_0)/(a_0-x_1)$, which again reduces the equation to
hypergeometric form. The solution is (specialising to $\kappa=6$
for percolation)
\begin{equation}\label{eqn:crossing}
P={\Gamma(\ffrac23)\over\Gamma(\ffrac43)\Gamma(\ffrac13)}\eta^{1/3}
{}_2F_1(\ffrac13,\ffrac23,\ffrac43;\eta)\,.
\end{equation}
It should be mentioned that this is but one of a number of
percolation crossing formulae. Another, conjectured by
Watts\cite{Watts}, for the probability that there is cluster which
simultaneously connects $AB$ to $CD$ and $BC$ to $DA$, has since
been proved by Dub\'edat\cite{Dubedat}. However, other formulae,
for example for the mean number of distinct clusters connecting
$AB$ and $CD$\cite{JCmean}, and for the probability that exactly
$N$ distinct clusters cross an annulus\cite{JCannulus}, are as yet
unproven using SLE methods.
\subsection{Critical exponents from SLE}\label{sec:sleexponents}
Many of the critical exponents which have previously been
conjectured by Coulomb gas or CFT methods may be derived
rigorously using SLE, once the underlying postulates are assumed
or proved. However SLE describes the measure on just a single
curve, and in the papers of LSW a great deal of ingenuity has gone
into showing how to relate this to all the other exponents. There
is not space in this article to do these justice. Instead we
describe three examples which give the flavour of the arguments,
which initially may appear quite unconventional compared with the
more traditional approaches.

\subsubsection{The fractal dimension of
SLE}\label{sec:fractaldim} The fractal dimension of any
geometrical object embedded in the plane can be defined roughly as
follows: let $N(\epsilon)$ be the minimum number of small discs of
radius $\epsilon$ required to cover the object. Then if
$N(\epsilon)\sim\epsilon^{-d_f}$ as $\epsilon\to0$, $d_f$ is the
fractal dimension.

One way of computing $d_f$ for a random curve $\gamma$ in the
plane is to ask for the probability $P(x,y,\epsilon)$ that a given
point $\zeta=x+iy$ lies within a distance $\epsilon$ of $\gamma$.
A simple scaling argument shows that if $P$ behaves like
$\epsilon^\delta f(x,y)$ as $\epsilon\to0$, then $\delta=2-d_f$.
We can derive a differential equation for $P$ along the lines of
the previous calculation. The only difference is that under the
conformal mapping $g_{dt}$, $\epsilon\to
|g_{dt}'(\zeta)|\epsilon\sim\big(1-2dt\,{\rm
Re}(1/\zeta^2)\big)\epsilon$. The differential equation (written
for convenience in cartesian coordinates) is
\begin{equation}\label{eqn:pde3}
\left({2x\over x^2+y^2}{\partial\over\partial x}-{2y\over
x^2+y^2}{\partial\over\partial y} +\frac\kappa
2{\partial^2\over\partial x^2}-{2(x^2-y^2)\over(x^2+y^2)^2}
\epsilon{\partial\over\partial\epsilon}\right)P=0\,.
\end{equation}
Now $P$ is dimensionless and therefore should have the form
$(\epsilon/r)^{2-d_f}$ times a function of the polar angle
$\theta$. In fact, the simple ansatz
$P=\epsilon^{2-d_f}y^\alpha(x^2+y^2)^\beta$, with $\alpha+2\beta=
d_f-2$ satisfies the equation. [The reason this works is connected
with the simple form for the correlator
$\langle\Phi_2\phi_{2,1}\phi_{2,1}\rangle$ discussed in
Sec.~\ref{sec:CS}.] This gives $\alpha=(\kappa-8)^2/8\kappa$,
$\beta=(\kappa-8)/2\kappa$ and
\begin{equation}
d_f=1+\kappa/8\,.
\end{equation}
This is correct for $\kappa\leq8$: otherwise there is another
solution with $\alpha=\beta=0$ and $d_f=2$. A more careful
statement and proof of this result can be found in \cite{Beffara}.

We see that the fractal dimension increases steadily from the
value 1 when $\kappa=0$ (corresponding to a straight line) to a
maximum value of 2 when $\kappa=8$. Beyond this value $\gamma$
becomes space-filling: every point in the upper half plane lies on
the curve.
\subsubsection{Crossing exponent}\label{sec:crossingexponent}
Consider a critical percolation problem in the upper half plane.
What is the asymptotic behaviour as $r\to\infty$ of the
probability that the interval $(0,1)$ on the real axis is
connected to the interval $(r,\infty)$? We expect this to decay as
some power of $r$. The value of this exponent may be found by
taking the appropriate limit of the crossing formula
(\ref{eqn:crossing}), but instead we shall compute it directly. In
order for there to be a crossing cluster, there must be two
cluster boundaries which also cross between the two intervals, and
which bound this cluster above and below. Denote the upper
boundary by $\gamma$. Then we need to know the probability $P(r)$
of there being another spanning curve lying between $\gamma$ and
$(1,r)$, averaged over all realisations of $\gamma$. Because of
the locality property, the measure on $\gamma$ is independent of
the existence of the lower boundary, and is given by SLE$_6$
conditioned not to hit the real axis along $(1,r)$. Note that
because $\kappa>4$ it will eventually hit the real axis at some
point to the right of $r$. For this reason we can do the
computation for general $\kappa>4$, although it gives the actual
crossing exponent only if $\kappa=6$.

Consider the behaviour of $P(r)$ under the conformal mapping $\hat
g_{dt}(z)\sim z+(2dt/z)-\sqrt\kappa dB_t$ (which maps the growing
tip $\tau_t$ into $0$.) The crossing probability should be
conformally invariant and depend only on the ratio of the lengths
of the two intervals, hence, by an argument which by now should be
familiar,
\begin{equation}
P(r)=\langle P\big(\hat g_{dt}(r)/\hat g_{dt}(1)\big)\rangle\,.
\end{equation}
Expanding this out, remembering as usual that $(dB_t)^2=dt$, and
setting to zero the $O(dt)$ term, we find for $r\gg1$
\begin{equation}
(\kappa-2)rP'(r)+\ffrac12\kappa r^2P''(r)=0\,,
\end{equation}
with the solution $P(r)\propto r^{-(\kappa-4)/\kappa}$ for
$\kappa>4$. Setting $\kappa=6$ then gives the result $\frac13$.

\subsubsection{The one-arm exponent}\label{sec:onearm}
Consider critical lattice percolation inside some finite region
(for example a disc of radius $R$). What is the probability that a
given site (e.g.~the origin) is connected to a finite segment $S$
of the boundary? This should decay like $R^{-\lambda}$, where
$\lambda$ is sometimes called the one-arm exponent. If we try to
formulate this in the continuum, we immediately run up against the
problem that all clusters are fractal with dimension $<2$, and so
the probability of any given point being in any given cluster is
zero. Instead, one may ask about the probability $P(r)$ that the
cluster connected to $S$ gets within a distance $r$ of the origin.
This should behave like $(r/R)^{\lambda}$. We can now set $R=1$
and treat the problem using radial SLE$_6$.

Consider now a radial SLE$_\kappa$ which starts at
$e^{i\theta_0}$. If $\kappa>4$ it will continually hit the
boundary. Let $P(\theta-\theta_0,t)$ be the probability that the
segment $(\theta_0,\theta)$ of the boundary has not been swallowed
by time $t$. Then, by considering the evolution as usual under
$g_{dt}$,
\begin{equation}
P(\theta,\theta_0,t)=\langle
P(\theta+d\theta,\theta_0+d\theta_0,t-dt)\rangle\,,
\end{equation}
where $d\theta=\cot\big((\theta-\theta_0)/2\big)dt$ and
$d\theta_0=\sqrt\kappa dB_t$. Setting $\theta_0=0$ and equating to
zero the $O(dt)$ term, we find the time-dependent differential
equation
\begin{equation}\label{eqn:pde4}
\partial_tP=\cot(\theta/2)\partial_\theta P+
\ffrac12\kappa\partial^2_\theta P\,.
\end{equation}
This has the form of a backwards Fokker-Plank equation.

Now, since $g_t'(0)=e^t$, it is reasonable that, after time $t$,
the SLE gets within a distance $O(e^{-t})$ of the origin.
Therefore we can interpret $P$ as roughly the probability that the
cluster connected to $(0,\theta)$ gets within a distance $r\sim
e^{-t}$ of the origin. A more careful argument \cite{LSW1arm}
confirms this. The boundary conditions are $P(0,t)=0$ as
$\theta\to0$, and (with more difficulty) $\partial_\theta
P(\theta,t)=0$ at $\theta=2\pi$. The solution may then be found by
inspection to be
\begin{equation}
P\propto e^{-\lambda t}\big(\sin(\theta/4)\big)^{1-4/\kappa}\,,
\end{equation}
where $\lambda=(\kappa^2-16)/32\kappa$. For percolation this gives
$\frac5{48}$, in agreement with Coulomb gas arguments
\cite{Nienhuis}.

The appearance of differential operators such as that in
(\ref{eqn:pde4}) will become clear from the CFT perspective in
Sec.~\ref{sec:CS}. If instead of choosing Neuman boundary
conditions at $\theta=2\pi$ we impose $P=0$, the same equation
gives the bulk 2-leg exponent $x_2$, which is also related to the
fractal dimension by $d_f=2-x_2$.
\newpage
\section{Relation to conformal field theory}\label{sec:cft}
\subsection{Basics of CFT.}\label{sec:basics}
The reader who already knows a little about CFT will have
recognised the differential equations in Sec.~\ref{sec:calc} as
being very similar in form to the BPZ equations\cite{BPZ}
satisfied by the correlation functions of a $\phi_{2,1}$ operator,
corresponding to a highest weight representation of the Virasoro
algebra with a level 2 null state.

For those readers for whom the above paragraph makes no sense, and
in any case to make the argument self-contained, we first
introduce the concepts of (boundary) conformal field theory
(BCFT.) We stress that these are heuristic in nature -- they serve
only as a guide to formulating the basic principles of CFT which
can then be developed into a mathematically consistent theory. For
a longer introduction to BCFT see \cite{JCBCFT} and, for a
complete account of CFT, \cite{YellowBook}.

We have at the back of our minds a euclidean field theory defined
as a path integral over some set of fundamental fields
$\{\psi(r)\}$. The partition function is $Z=\int
e^{-S(\{\psi\})}[d\psi]$ where the action $S(\{\psi\})=\int_{\cal
D}{\cal L}(\{\psi\})d^2\!r$ is an integral over a local lagrangian
density. These fields may be thought of as smeared-out continuum
versions of the lattice degrees of freedom. As in any field
theory, this continuum limit involves renormalisation. There are
so-called local scaling operators $\phi_j^{(0)}(r)$ which are
particular functionals of the fundamental degrees of freedom,
which have the property that we can define renormalised scaling
operators $\phi_j(r)=a^{-x_j}\phi_j^{(0)}(r)$ whose correlators
are finite in the continuum limit $a\to0$, that is
\begin{equation}
\lim_{a\to0}a^{-\sum_jx_j}\langle\phi_1^{(0)}(r_1)\ldots
\phi_N^{(0)}(r_N)\rangle=\langle\phi_1(r_1)\ldots
\phi_N(r_N)\rangle
\end{equation}
exists. The numbers $x_j$ are called the scaling dimensions, and
are related to the various critical exponents. They are related to
the conformal weights $(h_j,\bar h_j)$ by $x_j=h_j+\bar h_j$; the
difference $h_j-\bar h_j=s_j$ is called the spin of $\phi_j$, and
describes its behaviour under rotations. There are also \em
boundary operators\em, localised on the boundary, which have only
a single conformal weight equal to their scaling dimension.

The theory is developed independently of any particular set of
fundamental fields or lagrangian. An important role in this is
played by the stress tensor $T^{\mu\nu}(r)$, defined as the local
response of the action to a change in the metric:
\begin{equation}
\delta S=(1/4\pi)\int_{\cal D}T^{\mu\nu}\delta g_{\mu\nu}d^2r\,.
\end{equation}
Invariance under local rotations and scale transformations usually
implies that $T^{\mu\nu}$ is symmetric and traceless:
$T^\mu_\mu=0$. This also implies invariance under conformal
transformations, corresponding to $\delta g_{\mu\nu}\propto
f(r)g_{\mu\nu}$.

In two dimensional flat space, infinitesimal coordinate
transformations $r^\mu\to {r'}^\mu=r^\mu+\alpha^\mu(r)$ correspond
to infinitesimal transformations of the metric with $\delta
g^{\mu\nu}=-(\partial^\mu\alpha^\mu+
\partial^\nu\alpha^\nu)$. It is important to
realise that under these transformations the underlying lattice,
or UV cut-off, is not transformed. Otherwise they would amount to
a trivial reparametrisation. For a conformal transformation,
$\alpha^\mu(r)$ is given by an analytic function: in complex
coordinates $(z,\bar z)$, $\partial_{\bar z}\alpha^z =0$, so
$\alpha^z\equiv\alpha(z)$ is holomorphic. However, such a function
cannot be small everywhere (unless it is constant), so it is
necessary to consider coordinate transformations which are not
everywhere conformal.

Consider therefore two concentric semicircles $\Gamma_1$ and $\Gamma_2$
in the upper half plane, centred on the origin, and of radii $R_1<R_2$.
For $|r|<R_1$ let $\alpha^\mu$ be conformal, with $\alpha^z=\alpha(z)$,
while for $|r|>R_2$ take $\alpha^\mu=0$. In between,
$\alpha^\mu$ is not conformal, but is differentiable,
so that
$\delta S=(-1/2\pi)\int_{R_1<|r|<R_2}T^{\mu\nu}\partial_\mu\alpha_\nu d^2\!r$.
This can be integrated by parts to give a term
$(1/2\pi)\int_{R_1<|r|<R_2}\partial_\mu T^{\mu\nu}\alpha_\nu d^2\!r$ (which
must vanish because $\alpha_\nu$ is arbitrary in this region, implying
that $\partial_\mu T^{\mu\nu}=0$) and two surface terms. That on $\Gamma_2$
vanishes because $\alpha^\mu=0$ there. We are left with
\begin{equation}\label{eqn:deltaS}
\delta S=(1/2\pi)\int_{\Gamma_1}T^{\mu\nu}\alpha_\mu
\epsilon^{\nu\lambda}d\ell_\lambda\,,
\end{equation}
where $d\ell^\lambda$ is the line element along $\Gamma_1$.

The fact that $T^{\mu\nu}$ is conserved means, in complex
coordinates, that $\partial_{\bar z}T_{zz}=\partial_zT_{\bar z\bar
z}=0$, so that the correlations functions of $T(z)\equiv T_{zz}$
are holomorphic functions of $z$, while those of $\overline
T\equiv T_{\bar z\bar z}$ are antiholomorphic. (\ref{eqn:deltaS})
may then be written
\begin{equation}\label{eqn:deltaS2}
\delta S=(1/2\pi i)\int_{\Gamma_1}T(z)\alpha(z)dz+{\rm c.c.}\,.
\end{equation}

In any field theory with a boundary, it is necessary to impose
some boundary condition. It can be argued that any translationally
invariant boundary condition flows under the RG to conditions
satisfying $T_{xy}=0$, which in complex coordinates means that
$T=\overline T$ on the real axis. This means that the correlators
of $\overline T$ are those of $T$ analytically continued into the
lower half plane. The second term in (\ref{eqn:deltaS2}) may then
be dropped if the contour in the first term is around a complete
circle.

The conclusion of all this is that the effect of an infinitesimal
conformal transformation on any correlator of observables inside
$\Gamma_1$ is the same as inserting a contour integral $\int
T(z)\alpha(z)dz/2\pi i$ into the correlator.

Another important element of CFT is the operator product expansion
(OPE) of the stress tensor with other local operators. Since $T$
is holomorphic, this has the form
\begin{equation}\label{eqn:Tphi}
T(z)\cdot\phi(0)=\sum_n z^{-n-2}\phi^{(n)}(0)\,,
\end{equation}
where the $\phi^{(n)}$ are (possibly new) local operators. By
taking $\alpha(z)\propto z$ (corresponding to a scale
transformation) we see that $\phi^{(0)}=h\phi$, where $h$ is its
scaling dimension. Similarly, by taking $\alpha=$ const.,
$\phi^{(-1)}=\partial_x\phi$. Local operators for which
$\phi^{(n)}$ vanishes for $n\geq1$ are called primary. $T$ itself
is not primary: its OPE with itself takes the form
\begin{equation}\label{eqn:TT}
T(z)\cdot T(0)=c/2z^4+(2/z^2)T(0)+(1/z)\partial_zT(0)+\cdots\,,
\end{equation}
where $c$ is the conformal anomaly number, ubiquitous in CFT. For
example, the partition function on a long cylinder of length $L$
and circumference $\ell$ behaves as $\exp(\pi cL/\ell)$,
cf.~(\ref{eqn:c6}).
\subsection{Radial quantisation}
This is the most important concept in understanding the link
between SLE and CFT. We introduce it in the context of boundary
CFT. As before, suppose there is some set of fundamental fields
$\{\psi(r)\}$, with a Gibbs measure $e^{-S[ \psi]}[d\psi]$. Let
$\Gamma$ be a semicircle in the upper half plane, centered on the
origin. The Hilbert space of the BCFT is the function space (with
a suitable norm) of field configurations $\{\psi_\Gamma\}$ on
$\Gamma$.

The vacuum state is given by weighting each state
$|\psi'_\Gamma\rangle$ by the (normalised) path integral
restricted to the interior of $\Gamma$ and conditioned to take the
specified values $\psi'_\Gamma$ on the boundary:
\begin{equation}
|0\rangle=\int[d\psi'_\Gamma]\int_{\psi_\Gamma=\psi'_\Gamma}
[d\psi]\,e^{-S[\psi]}\,|\psi'_\Gamma\rangle\,.
\end{equation}
Note that because of scale invariance different choices of the
radius of $\Gamma$ are equivalent, up to a normalisation factor.

Similarly, inserting a local operator $\phi(0)$ at the origin into
the path integral defines a state $|\phi\rangle$. This is called
the operator-state correspondence of CFT. If we also insert
$(1/2\pi i)\int_Cz^{n+1}T(z)dz$, where $C$ lies inside $\Gamma$,
we get a state $L_n|\phi\rangle$. The $L_n$ act linearly on the
Hilbert space. From the OPE (\ref{eqn:Tphi}) we see that
$L_n|\phi\rangle\propto|\phi^{(n)}\rangle$, and that, in
particular, $L_0|\phi\rangle=h_\phi|\phi\rangle$. If $\phi$ is
primary, $L_n|\phi\rangle=0$ for $n\geq1$. From the OPE
(\ref{eqn:TT}) of $T$ with itself follow the commutation relations
for the $L_n$
\begin{equation}\label{eqn:vir}
[L_n,L_m]=(n-m)L_{n+m}+\ffrac1{12}cn(n^2-1)\delta_{n,-m}\,,
\end{equation}
which are known as the Virasoro algebra. The state $|\phi\rangle$
together with all its descendants, formed by acting on
$|\phi\rangle$ an arbitrary number of times with the $L_n$ with
$n\leq-1$, give a highest weight representation (where the weight
is defined as the eigenvalue of $-L_0$.)

There is another way of generating such a highest weight
representation. Suppose the boundary conditions on the negative
and positive real axes are both conformal, that is they satisfy
$T=\overline T$, but they are different. The vacuum with these
boundary conditions gives a highest weight state which it is
sometimes useful to think of as corresponding to the insertion of
a `boundary condition changing' (bcc) operator at the origin. An
example is the continuum limit of an Ising model in which the
spins on the negative real axis are $-1$, and those on the
positive axis are $+1$.

\subsection{Curves and states.}\label{sec:curves} In this section
we describe a way of associating states in the Hilbert space of
the BCFT with the growing curves of the Loewner process. This was
first understood by M.~Bauer and D.~Bernard\cite{BB}, but we shall
present the arguments slightly differently.

The boundary conditions associated with a bcc operator guarantee
the existence, on the lattice, of a domain wall connecting the
origin to infinity. Given a particular realisation $\gamma$, we
can condition the Ising spins on its existence. We would like to
be able to assume that this property continues to hold in the
continuum limit: that is, we can condition the fields $\{\psi\}$
on the existence of a such a curve. However, this involves
conditioning on an event with probability zero: it turns out that
in general the probability that, with respect to the measure in
the path integral, the probability that a domain wall hits the
real axis somewhere in an interval of length $\epsilon$ vanishes
like $\epsilon^h$. In what follows we shall regard $\epsilon$ as
small but fixed, and assume that the usual properties of SLE are
applicable to this more general case.

Any such curve may be generated by a Loewner process: denote as
before the part of the curve up to time $t$ by $\gamma_t$. The
existence of this curve depends on only the field configurations
$\psi$ in the interior of $\Gamma$, as long as $\gamma_t$ lies
wholly inside this region. Then we can condition the fields
contributing to the path integral on the existence of $\gamma_t$,
thus defining a state
\begin{equation}
|\gamma_t\rangle =
\int[d\psi'_\Gamma]\int_{\psi_\Gamma=\psi'_\Gamma;\gamma_t}
[d\psi]\,e^{-S[\psi]}\,|\psi'_\Gamma\rangle\,.
\end{equation}
The path integral (over the whole of the upper half plane, not
just the interior of $\Gamma$), when conditioned on $\gamma_t$,
gives a measure $d\mu(\gamma_t)$ on these curves. The state
\begin{equation}\label{eqn:ht}
|h\rangle=|h_t\rangle\equiv\int d\mu(\gamma_t)|\gamma_t\rangle
\end{equation}
is in fact independent of $t$, since it is just given by the path
integral conditioned on there being a curve connecting the origin
to infinity, as guaranteed by the boundary conditions. In fact we
see that $|h\rangle$ is just the state corresponding to a boundary
condition changing operator at the origin.

However, $d\mu(\gamma_t)$ is also given by the measure on $a_t$ in
Loewner evolution, through the iterated sequence of conformal
mappings satisfying $d\hat g_t=2dt/\hat g_t-da_t$. This
corresponds to an infinitesimal conformal mapping of the upper
half plane minus $K_t$. As explained in the previous section,
$d\hat g_t$ corresponds to inserting $(1/2\pi
i)\int_C(2dt/z-da_t)T(z)dz$. In operator language, this
corresponds to acting on $|\gamma_t\rangle$ with
$2L_{-2}dt-L_{-1}da_t$ where $L_n=(1/2\pi i)\int_Cz^{n+1}T(z)dz$.
Thus, for any $t_1<t$,
\begin{equation}
|g_{t_1}(\gamma_t)\rangle ={\bf
T}\exp\left(\int_0^{t_1}\big(2L_{-2}dt'-L_{-1}da_{t'}\big)\right)
|\gamma_t\rangle\,,
\end{equation}
where $\bf T$ denotes a time-ordered exponential.

The measure on $\gamma_t$ is the product of the measure of
$\gamma_t\setminus\gamma_{t_1}$, conditioned on $\gamma_{t_1}$,
with the unconditioned measure on $\gamma_{t_1}$. The first is the
same as the unconditioned measure on $g_{t_1}(\gamma_t)$, and the
second is given by the measure on $a_{t'}$ for $t'\in[0,t_1]$.
Thus we can rewrite both the measure and the state in
(\ref{eqn:ht}) as
\begin{equation}
|h_t\rangle=\int d\mu(g_{t_1}(\gamma_t))\int
d\mu(a_{t';t'\in[0,t_1]}) {\bf
T}e^{\int_{t_1}^0\big(2L_{-2}dt'-L_{-1}da_{t'}\big)}
|g_{t_1}(\gamma_t)\rangle\,.
\end{equation}
For SLE, $a_t$ is proportional to a Brownian process. The
integration over realisations of this for $t'\in[0,t_1]$ may be
performed by breaking up the time interval into small segments of
size $\delta t$, expanding out the exponential to $O(\delta t)$,
using $(B_{\delta t})^2\approx \delta t$, and re-exponentiating.
The result is
\begin{equation}
|h_t\rangle=\exp\left(-\big(2L_{-2}-(\kappa/2)L_{-1}^2\big)t_1\right)
|h_{t-t_1}\rangle\,.
\end{equation}
But, as we argued earlier, $|h_t\rangle$ is independent of $t$,
and therefore
\begin{equation}
\label{l2l1} \big(2L_{-2}-(\kappa/2)L_{-1}^2\big)|h\rangle=0\,.
\end{equation}

This means that the descendant states $L_{-2}|h\rangle$ and
$L^2_{-1}|h\rangle$ are linearly dependent. We say that the
Virasoro representation corresponding to $|h\rangle$ has a null
state at level 2. From this follow an number of important
consequences. Acting on (\ref{l2l1}) with $L_1$ and $L_2$, and
using the Virasoro algebra (\ref{eqn:vir}) and the fact that
$L_1|h\rangle=L_2|h\rangle=0$ while $L_0|h\rangle=h|h\rangle$,
leads to
\begin{eqnarray}
h&=&h_{2,1}={6-\kappa\over2\kappa}\,;\\
c&=&{(3\kappa-8)(6-\kappa)\over2\kappa}\,.
\end{eqnarray}
These are the fundamental relations between the parameter $\kappa$
of SLE and the data of CFT. The conventional notation $h_{2,1}$
comes from the Kac formula in CFT which we do not need here. In
fact this is appropriate to the case $\kappa<4$: for $\kappa>4$ it
corresponds to $h_{1,2}$. (To further confuse the matter, some
authors reverse the labels.) Note that the boundary exponent $h$
parametrises the failure of locality in (\ref{eqn:kappa6}). From
CFT we may also deduce that, with respect to the path integral
measure, the probability that a curve connects small intervals of
size $\epsilon$ about points $r_1$, $r_2$ on the real axis behaves
like
\begin{equation}
\epsilon^{2h_{2,1}}\langle\phi_{2,1}(r_1)\phi_{2,1}(r_2)\rangle
\propto\left({\epsilon\over|r_1-r_2|}\right)^{2h_{2,1}}\,.
\end{equation}
Such a result, elementary in CFT, is difficult to obtain directly
from SLE in which the curves are conditioned to begin and end at
\em given \em points.

Note that the central charge $c$ vanishes when either locality
($\kappa=6$) or restriction ($\kappa=\frac83$) hold. These cases
correspond to the continuum limit of percolation and self-avoiding
walks respectively, corresponding to formal limits $Q\to1$ in the
Potts model and $n\to0$ in the $O(n)$ model for which the
unconditioned partition function is trivial.

\subsection{Differential Equations}
In this section we discuss how the linear second order
differential equations for various observables which arise from
the stochastic aspect of SLE follow equivalently from the null
state condition in CFT. In this context they are known as the BPZ
equations\cite{BPZ}. As an example consider Schramm's formula
(\ref{eqn:schrammformula}) for the probability $P$ that a point
$\zeta$ lies to the right of $\gamma$, or equivalently the
expectation value of the indicator function ${\cal O}(\zeta)$
which is 1 if this is satisfied and zero otherwise. In SLE, this
expectation value is with respect to the measure on curves which
connect the point $a_0$  to infinity. In CFT, as explained above,
we can only consider curves which intersect some
$\epsilon$-neighbourhood on the real axis. Therefore $P$ should be
written as a ratio of expectation values with respect to the CFT
measure:
\begin{equation}
P(\zeta;a_0)=\lim_{r_2\to\infty}{\langle\phi_{2,1}(a_0){\cal
O}(\zeta)\phi_{2,1}(r_2)\rangle\over
\langle\phi_{2,1}(a_0)\phi_{2,1}(r_2)\rangle}\,.
\end{equation}
We can derive differential equations for the correlators in the
numerator and denominator by inserting into each of them a factor
$(1/(2\pi i)\int_\Gamma\alpha(z)T(z)dz+$ c.c., where
$\alpha(z)=2/(z-a_0)$, and $\Gamma$ is a small semicircle
surrounding $a_0$. This is equivalent to making the infinitesimal
transformation $z\to z+2\epsilon/(z-a_0)$. As before, the
c.c.~term is equivalent to extending the contour in the first term
to a full circle. The effect of this insertion may be evaluated in
two ways: by shrinking the contour onto $a_0$ and using the OPE
between $T$ and $\phi_{2,1}$ we get
\begin{equation}\label{var1}
2L_{-2}\phi_{2,1}(a_0)=(\kappa/2)L_{-1}^2\phi_{2,1}(a_0)=
\partial^2_{a_0}\phi_{2,1}(a_0)\,,
\end{equation}
while wrapping it about $\zeta$ (in a clockwise sense) we get
\begin{equation}\label{var2}
-\big(2/(\zeta-a_0)\big)\partial_\zeta{\cal
O}-\big(2/(\bar\zeta-a_0)\big)\partial_{\bar\zeta}{\cal O}\,.
\end{equation}
The effect on $\phi_{2,1}(r_2)$ vanishes in the limit
$r_2\to\infty$. As a result we can ignore the variation of the
denominator in this case. Equating (\ref{var1}) and (\ref{var2})
inside the correlation function in the numerator then leads to the
differential equation (\ref{eqn:pde1}) for $P$ found in
Sec.~\ref{sec:schrammformula}.

\subsubsection{Calagero-Sutherland model}\label{sec:CS}
While many of the results of SLE may be re-derived in CFT with less
rigour but perhaps greater simplicity, the latter contains much
more information which is not immediately apparent from the SLE
perspective. For example, one may consider correlation functions
$\langle\phi_{1,2}(r_1)\phi_{1,2}(r_2)
\ldots\phi_{2,1}(r_N)\ldots\rangle$ of multiple boundary condition
changing operators with other operators either in the bulk or on
the boundary. Evaluating the effect of an insertion $(1/2\pi
i)\int_\Gamma T(z)dz/(z-r_j)$ where $\Gamma$ surrounds $r_j$ leads
to a second order differential equation satisfied by the
correlation function for \em each \em $j$.

This property is very powerful in the radial version. Consider the
correlation function
\begin{equation}
C_\Phi(\theta_1,\ldots,\theta_N)=\langle\phi_{2,1}(\theta_1)
\ldots\phi_{2,1}(\theta_N)\Phi(0)\rangle
\end{equation}
of $N$ $\phi_{2,1}$ operators on the boundary of the unit disc
with a single bulk operator $\Phi$ at the origin. Consider the
effect of inserting $(1/2 \pi i)\int_\Gamma\alpha_j(z)T(z)dz$
where (cf. (\ref{eqn:radialsle}))
\begin{equation}
\alpha_j(z)=-z{z+e^{i\theta_j}\over z-e^{i\theta_j}}
\end{equation}
and $\Gamma$ surrounds the origin. Once again, this may be
evaluated in two ways: by taking $\Gamma$ up to the boundary, with
exception of small semicircles around the points $e^{i\theta_k}$,
we get $G_jC_\Phi$, where $G_j$ is the second order differential
operator
\begin{equation}
G_j=-\frac\kappa
2{\partial^2\over\partial\theta_j^2}+\frac{h_{2,1}}6+\frac c{12} -
\sum_{k\not=j}\left(\cot{\theta_k-\theta_j\over2}
{\partial\over\partial\theta_k}
-{1\over2\sin^2(\theta_k-\theta_j)/2}h_{2,1}\right)\,.
\end{equation}
The first three terms come from evaluating the contour integral
near $e^{i\theta_j}$, where $\alpha_j$ acts like
$2L_{-2}-\frac16L_0-\frac{c}{12}$, (the term $\frac{c}{12}$ comes
from the curvature of the boundary,) and the term with $k\not=j$
from the contour near $e^{i\theta_k}$, where it acts like
$\alpha_j(e^{i\theta_k})L_{-1}+{\rm
Re}\,\alpha_j'(e^{i\theta_k})L_0$.

On the other hand, shrinking the contour down on the origin we see
that $\alpha_j(z)=z+O(z^2)$, so that on $\Phi(0)$ it has the
effect of $L_0+\overline L_0+\ldots$, where the omitted terms
involve the $L_n$ and $\overline L_n$ with $n>0$. Assuming that
$\Phi$ is primary, these other terms vanish, leaving simply
$(L_0+\overline L_0)\Phi=x_\Phi\Phi$. Equating the two evaluations
we find the differential equation
\begin{equation}
G_j\,C_\Phi=x_\Phi\,C_\Phi\,.
\end{equation}

In general there is an $(N-1)$-dimensional space of independent
differential operators $G_j$ with a common eigenfunction $C_\Phi$.
(There is one fewer dimension because they all commute with the
total angular momentum $\sum_j(\partial/\partial\theta_j)$.) For
the case $N=2$, setting $\theta=\theta_2-\theta_1$, we recognise
the differential operator in Sec.~\ref{sec:onearm}.

In general these operators are not self-adjoint and their spectrum
is difficult to analyse. However, if we form the equally-weighted
linear combination $G\equiv \sum_{j=1}^NG_j$, the terms with a
single derivative may be written in the form $\sum_k (\partial
V/\partial\theta_k)(\partial/\partial\theta_k)$ where $V$ is a
potential function. In this case it is well known from the theory
of the Fokker-Plank equation that $G$ is related by a similarity
transformation to a self-adjoint operator. In fact\cite{JCCS} if
we form $|\Psi_N|^{2/\kappa}G|\Psi_N|^{-2/\kappa}$ where $\Psi_N=
\prod_{j<k}\big(e^{i\theta_j}-e^{\theta_k}\big)$ is the
`free-fermion' wave function on the circle, the result is, up to
calculable constants the well-known $N$-particle
Calogero-Sutherland hamiltonian
\begin{equation}
H_N(\beta)=-\frac12\sum_{j=1}^N{\partial^2\over\partial\theta_j^2}
+{\beta(\beta-2)\over16}\sum_{j<k}{1\over\sin^2(\theta_j-\theta_k)/2}\,,
\end{equation}
with $\beta=8/\kappa$. It follows that the scaling dimensions of
bulk operators like $\Phi$ are simply related to eigenvalues
$\Lambda_N$ of $H_N$ by
\begin{equation}
x_\Phi=(\kappa/N)\Lambda_N(8/\kappa)-(4/\kappa N)E_N^{\rm
ff}+\ffrac16h_{2,1}+\ffrac1{12}c\,,
\end{equation}
where $E_N^{\rm ff}=\frac1{24}N(N^2-1)$. Similarly $C_\Phi$ is
proportional to the corresponding eigenfunction. In fact the
ground state (with conventional boundary conditions) turns out to
correspond to the bulk $N$-leg operator discussed in
Sec.~\ref{Nleg1}. The corresponding correlator is
$|\Psi_N|^{2/\kappa}$.

\newpage
\section{Related ideas}\label{sec:relatedideas}
\subsubsection{Multiple SLEs}\label{sec:multiplesles}
We pointed out earlier that the boundary operators $\phi_{2,1}$
correspond to the continuum limits of lattice curves which hit
the boundary at a given point. For a single curve, these are
described by SLE, and we have shown in that case how the resulting
differential equations also appear in CFT. Using the $N$-particle
generalisation of the CFT results of the previous section, we may
now `reverse engineer' the problem and conjecture the
generalisation of SLE to $N$ curves.

The expectation value of some observable $\cal O$ given that $N$
curves, starting at the origin, hit the boundary at
$(\theta_1,\ldots,\theta_N)$ is
\begin{equation}
P_{\cal O}(\theta_1,\ldots,\theta_N) = {F_{\cal
O}(\theta_1,\ldots,\theta_N)\over F_{\bf
1}(\theta_1,\ldots,\theta_N)}\,,
\end{equation}
where $F_{\cal O}=\langle{\cal O}\,\phi_{2,1}(e^{i\theta_1})
\ldots \phi_{2,1}(e^{i\theta_N})\Phi_N(0)\rangle$. This satisfies
the BPZ equation
\begin{equation}
G_j\,F_{\cal O}=\langle(\delta_j{\cal O}\phi_{2,1}(e^{i\theta_1})
\ldots \phi_{2,1}(e^{i\theta_N})\Phi_N(0)\rangle\,,
\end{equation}
where $\delta_j{\cal O}$ is the variation in $\cal O$ under
$\alpha_j$. If we now write $F_{\cal O}=F_{\bf 1}\cdot P_{\cal O}$
and use the fact that $G_jF_{\bf 1}=x_\Phi F_{\bf 1}$, we find a
relatively simple differential equation for $P_{\cal O}$, since
the non-derivative terms in $G_j$ cancel. There is also a
complication since the second derivative gives a cross term
proportional to $\big(\partial_{\theta_j}F_{\bf 1}\big)
\big(\partial_{\theta_j}P_{\cal O}\big)$. However, this may be
evaluated from the explicit form $F_{\bf 1}=|\Psi_N|^{2/\kappa}$.
The result is
\begin{equation}
\left(\frac\kappa2{\partial^2\over\partial\theta_j^2} +
\sum_{k\not=j}\cot{\theta_k-\theta_j\over2}
\left({\partial\over\partial\theta_k}-
{\partial\over\partial\theta_j}\right)\right)P_{\cal O}=
\delta_jP_{\cal O}\,,
\end{equation}
where the right hand side comes from the variation in $\cal O$.

The left hand side may be recognised as the generator (the adjoint
of the Fokker-Planck operator) for the stochastic process
\begin{eqnarray}
d\theta_j&=&\sqrt\kappa
dB_t+\sum_{k\not=j}(\rho_k/2)\cot\big((\theta_j-\theta_k)/2\big)\,
dt\,;\label{mult1}\\
d\theta_k&=&\cot\big((\theta_k-\theta_j)/2\big)\,dt\,,\label{mult2}
\end{eqnarray}
where $\rho_k=2$. [For general values of the parameters $\rho_k$
this process is known as (radial) SLE$(\kappa,\vec\rho\,)$,
although this is more usually considered in the chordal version.
It has been argued\cite{JCSLEkr} that this applies to the level
lines of a free gaussian field with piecewise constant Dirichlet
boundary conditions: the parameters $\rho_k$ are related to the
size of the discontinuities at the points $e^{i\theta_k}$.
SLE$(\kappa,\vec\rho\,)$ has also been used to give examples of
restriction measures on curves which are not reflection
symmetric\cite{SLEkrrefs}.]

We see that $e^{i\theta_j}$ undergoes Brownian motion but is also
repelled by the other particles at $e^{i\theta_k}$ ($k\not=j$):
these particles are themselves repelled deterministically from
$e^{i\theta_j}$. The infinitesimal transformation $\alpha_j$
corresponds to the radial Loewner equation
\begin{equation}\label{mult3}
{dg_{j,t}\over dt}=-g_{j,t}\,{g_{j,t}+e^{i\theta_{j,t}} \over
g_{j,t}-e^{i\theta_{j,t}}}\,.
\end{equation}

The conjectured interpretation of this is as follows: we have $N$
non-intersecting curves connecting the boundary points
$e^{i\theta_{k,0}}$ to the origin. The evolution of the $j$th
curve in the presence of the others is given by the radial Loewner
equation with, however, the driving term not being simple Brownian
motion but instead the more complicated process
(\ref{mult1},\ref{mult2}).

However, from the CFT point of view we may equally well consider
the linear combination $\sum_jG_j$. The Loewner equation is now
\begin{equation}\label{mult4}
\dot g_t=-g_t\sum_{j=1}^N{g_t+e^{i\theta_{j,t}} \over
g_t-e^{i\theta_{j,t}}}\,,
\end{equation}
where
\begin{equation}\label{mult5}
d\theta_j=\sqrt\kappa
dB_t^j+2\sum_{k\not=j}\cot\big((\theta_j-\theta_k)/2\big)\,dt\,.
\end{equation}
This is known in the theory of random matrices as Dyson's Brownian
motion. It describes the statistics of the eigenvalues of unitary
matrices. The conjectured interpretation is now in terms of $N$
random curves which are all growing in each other's mutual
presence at the \em same \em mean rate (measured in Loewner time).
From the point of view of SLE, it is by no means obvious that the
measure on $N$ curves generated by process
(\ref{mult1},\ref{mult2},\ref{mult3}) is the same as that given by
(\ref{mult4},\ref{mult5}). However CFT suggests that, for curves
which are the continuum limit of suitable lattice models, this is
indeed the case.
\subsection{Other variants of SLE}\label{sec:variants}
So far we have discussed only chordal SLE, which describes curves
connecting distinct points on the boundary of a simple connected
domain, and radial SLE, in which the curve connects a boundary
point to an interior point. Another simple variant is dipolar
SLE\cite{dipolar}, in which the curve is constrained to start at
boundary point and to end on some finite segment of the boundary
not containing the point. The canonical domain is an infinitely
long strip, with the curve starting a point on one edge and ending
on the other edge. This set-up allows the computation of several
interesting physical quantities.

The study of SLE in multiply-connected domains is very
interesting. Their conformal classes are characterised by a set of
moduli, which change as the curve grows. R.~Friedrich and
co-workers\cite{BFriedrich} have argued that SLE in such a domain
is characterised by diffusion in moduli space as well as diffusion
on the boundary.

It is possible to rewrite the differential equations which arise
from null state conditions in extended CFTs (for example
super-conformal CFTs\cite{Ras} and WZWN models\cite{GLW}) in terms
of the generators of stochastic conformal mappings which
generalise that of Loewner. However, a physical interpretation in
terms of the continuum limit of lattice curves appears so far to
be missing.

\subsection{Other growth models}
SLE is in fact just one very special, solvable, example of an
approach to growth processes in two dimensions using conformal
mappings which has been around for a number of years. For a recent
review see \cite{BBrev}. The prototypical problem of this type is
diffusion-limited aggregation (DLA). In this model of cluster
formation, particles of finite radius diffuse in, one by one, from
infinity until they hit the existing cluster, where they stick.
The probability of sticking at a given point is proportional to
the local electric field, if we imagine the cluster as being
charged. The resultant highly branched structures are very similar
to those observed in smoke particles, and in viscous fingering
experiments where one fluid is forced into another in which it is
immiscible. Hastings and Levitov\cite{HL} proposed an approach to
this problem using conformal mappings. At each time $t$, the
boundary of the cluster is described by the conformal mapping
$f_t(z)$ which takes it to the unit disc. The cluster is grown by
adding a small semicircular piece to the boundary. The way this
changes $f_t$ is well known according to a theorem of Hadamard.
The difficulty is that the probability of adding this piece at a
given point depends on the local electric field which itself
depends on $f_t'$. The equation of motion for $f_t$ is therefore
more complicated than in SLE. Moreover it may be shown that almost
all initially smooth boundary curves evolve towards a finite-time
singularity: this is thought to be responsible for branching, but
just at this point the equations must be regularised to reflect
the finite size of the particles (or, in viscous fingering, the
effects of finite surface tension.)

It is also possible to generate branching structures by making the
driving term $a_t$ in Loewner's equation discontinuous, for
example taking it to be a Levy process. Unfortunately this does
not appear to describe a physically interesting model.

Finally, Hastings\cite{Hastings} has proposed two related growth
models which each lead, in the continuum limit, to SLE. These are
very similar to DLA, except that growth is only allowed at the
tip. The first, called the arbitrary Laplacian random walk, takes
place on the lattice. The tip moves to one of the neighbouring
unoccupied sites $r$ with relative probability $E(r)^\eta$, where
$E(r)$ is the lattice electric field, that is the potential
difference between the tip and $r$, and $\eta$ is a parameter. The
second growth model takes place in the continuum via iterated
conformal mappings, in which pieces of length $\ell_1$ are added
to the tip, but shifted to the left or right relative to the
previous growth direction by a random amount $\pm\ell_2$. This
model depends on the ratio $\ell_2/\ell_1$, and leads, in the
continuum limit, to SLE$_\kappa$ with $\kappa=\ell_2/4\ell_1$. For
the lattice model there is no universal relation between $\kappa$
and $\eta$, except for $\eta=1$, which is the same as the
loop-erased random walk (Sec.~\ref{sec:Onmodel}) and converges to
SLE$_2$.

\end{document}